\documentclass[reprint,aps,prl, 
nobibnotes,
showkeys,
balancelastpage,
a4paper,floatfix,
superscriptaddress]{revtex4-2}
\usepackage{fix-cm}
\usepackage[margin=0.75in]{geometry}
\usepackage[T1]{fontenc}
\usepackage{graphicx}
\usepackage{amsmath,amssymb,bm}
\usepackage[cal=cm]{mathalfa}
\usepackage{xcolor}
\usepackage[normalem]{ulem}
\usepackage{booktabs}
\usepackage{enumitem}
\usepackage{natbib}

\usepackage{multirow}
\usepackage{dcolumn}

\usepackage{makecell}

\usepackage[rightcaption]{sidecap}
\usepackage{mathtools}
\usepackage{relsize}
\usepackage{newtxtext}
\usepackage{newtxmath}
\usepackage[scaled=0.86]{helvet}
\usepackage{graphicx}
\usepackage{subfigure}  
\usepackage[percent]{overpic}
\usepackage{tikz}
\usetikzlibrary{arrows.meta,calc,positioning}

\usepackage{appendix}
\usepackage{amsmath}
\usepackage{needspace}

\usepackage[framemethod=tikz]{mdframed}

\usepackage{hyperref}
\hypersetup{colorlinks=true,citecolor=[rgb]{0,0.18,0.68}, 
urlcolor=[rgb]{0,0.18,0.68}, linkcolor=[rgb]{0,0.18,0.68}, final = true}

\newmdenv[
  linewidth=0.6pt,
  roundcorner=4pt,
  linecolor=black,
  backgroundcolor=white,
  innertopmargin=6pt,
  innerbottommargin=6pt,
  innerleftmargin=6pt,
  innerrightmargin=6pt,
  skipabove=6pt,
  skipbelow=6pt
]{figframe}

\tikzset{
  nicearrow/.style={-{Latex[length=3mm]}, line width=0.9pt},
  notebox/.style={draw, rounded corners, fill=white, fill opacity=.85, text opacity=1,
                  inner sep=3pt, line width=0.5pt}
}

\usepackage{siunitx}
\DeclareSIUnit\angstrom{\text {Å}}

  \setcitestyle{super}

  \def\be{\begin{equation*}}
  \def\ee{\end{equation*}}
  \def\ba{\begin{eqnarray}}
  \def\ea{\end{eqnarray}}

  \def\sb#1{\textbf{\bt{#1}}}
  \def\nsb#1{\noindent\textbf{\bt{#1~}}}
  
  \definecolor{or}{RGB}{234,142,53}
  \definecolor{gr}{RGB}{150,150,150}
  \definecolor{bl}{RGB}{54,152,187}

  \newcommand{\ie}{\textit{i.e.}}
  \newcommand{\eg}{\textit{e.g.}}

  \def\sb#1{\textbf{\textsf{#1}}}
  \def\nsb#1{\noindent\textbf{\textsf{#1~}}}

  \definecolor{YKB}{rgb}{0.00,0.18,0.65}

\def\Label#1{}

\begin{document}

\title{%
  \texorpdfstring
    {\larger\sb{Geometric--Chemical Distance Between Protein Surfaces}}
    {Geometric--Chemical Distance Between Protein Surfaces}%
}

\author{Himanshu Swami}
\email{swamihimanshush@gmail.com}
\affiliation{Department of Physics, Ulsan National Institute of Science and Technology, Ulsan 44919, South Korea}
\author{John M. McBride}
\email{jmmcbride@protonmail.com}
\affiliation{Acoustics Research Institute, Austrian Academy of Sciences, Vienna, Austria}
\affiliation{Department of Physics, Ulsan National Institute of Science and Technology, Ulsan 44919, South Korea}
\author{Jean-Pierre Eckmann}\email{jean-pierre.eckmann@unige.ch}
\affiliation{D\'epartement de Physique Th\'eorique et Section de
  Math\'ematiques, Universit\'e de Gen\`eve, Geneva, Switzerland}
\author{Tsvi Tlusty}
\email{tsvitlusty@gmail.com}
\affiliation{Department of Physics, Ulsan National Institute of Science and Technology, Ulsan 44919, South Korea}

\begin{abstract}
\nsb{Abstract}\\
Proteins recognize, bind, and catalyze through molecular surfaces, where geometry and chemical patterning determine interaction. Comparing these surfaces requires both a geometric--chemical distance and a correspondence that relates one complete surface to another.

Here we introduce IFACE (Intrinsic Field--Aligned Coupled Embedding). IFACE derives a symmetric geometric--chemical distance by optimizing a probabilistic coupling over intrinsic geometry, mean curvature, electrostatics, hydrophobicity, and hydrogen-bond propensity. The same coupling provides an explicit surface map.

For molecular-dynamics conformers, IFACE distinguishes the same protein from distinct proteins more accurately than TM-distance and a
Laplace--Beltrami spectral distance. A Jensen--Shannon distribution distance performs best in this binary identity test, because aggregate surface-feature distributions already identify each protein.

A distance must also satisfy a global requirement: its pairwise values must place many distinct protein surfaces consistently in one space. We therefore tested IFACE across six protein families. It produces the strongest family classification and clustering among the distributional, spectral, MaSIF, and SurfaceID comparisons.

The inferred maps preserve geodesic neighborhoods and transfer
heme-centered pocket regions across cytochrome P450 proteins. IFACE therefore provides, from one construction, both a distance between complete protein surfaces and the local map that explains that distance.
\end{abstract}


\maketitle
\section{\sb{Introduction}}

Molecular recognition, catalysis, regulation, and assembly occur at protein surfaces~\cite{janin1995protein,hartwell1999molecular,goodsell2000structural,tlusty2025life,weinreb2025enzymes,mcbride2024physical}.
More broadly, protein form and function are shaped by the interplay among mechanics, chemistry and evolution~\cite{zeldovich2008,liberles2012,tlusty2016,tlusty2017physical,dutta2018green,eckmann2019}.
The relationship between sequence and folded structure is well established~\cite{anfinsen1973principles,chothia1986relation}, and modern sequence-based methods now predict folded structures with high accuracy~\cite{jumper2021highly,baek2021accurate}.
Recent work further indicates that these models can capture subtle mutation-induced structural responses~\cite{mcbride2023alphafold2,mcbride2024ai,arroyogomez2025,vanaalst2025}.
Yet a central problem remains: how to compare proteins at the level of their \emph{interfaces}~\cite{lawrence1993shape,shulman2004recognition,chartier2016}.
Functional specificity often resides at the exposed surface rather than in the global fold~\cite{jones1997analysis,glaser2003}.

A protein surface is a curved interface formed by solvent-exposed side chains~\cite{richards1977areas}, where geometry and chemistry jointly determine interaction. Curvature constrains approach; electrostatics, hydrophobicity, and hydrogen-bonding shape interaction specificity~\cite{sheinerman2000electrostatic,kyte1982simple,kortemme2003orientation,konc2007protein,bental2018}. These chemical features are themselves statistically correlated with surface curvature~\cite{mcbride2024}. The molecular boundary is not unique: common choices include the van der Waals surface, the solvent-accessible surface (SAS)~\cite{lee1971interpretation}, and the solvent-excluded surface (SES)~\cite{connolly1983analytical}. Here we use SES, which approximates the interface encountered by solvent and binding partners; the framework is not tied to this particular surface definition.

Despite the physical coupling of geometry and chemistry at protein surfaces, existing approaches typically compare surfaces through global shape, spectral geometry, descriptors, or learned local representations.  
In contrast, we formulate protein-surface comparison as a correspondence problem in which intrinsic geometry, extrinsic curvature, and spatially distributed chemical fields jointly determine the map, and a symmetric distance between the complete surfaces is derived from that map.

Protein-surface comparison has developed along two complementary lines. Classical methods compare shape, intrinsic geometry, or physicochemical descriptors, often through global summaries or surface projections~\cite{la20093d,yin2009fast,venkatraman2009protein,kihara2011molecular,hass2014round,zhu2015large,sirugue2024plo3s,schweke2022surfmap}. Learning-based methods, including MaSIF, DeeplyTough, DeepSurf, and SurfaceID, integrate geometric and physicochemical information in representations designed for interface recognition, binding-site detection, and local patch retrieval~\cite{gainza2020deciphering,simonovsky2020deeplytough,mylonas2021deepsurf,riahi2023surface,lin2024exploiting}. These approaches provide powerful measures of surface similarity, but do not generally yield both a coherent correspondence between complete surfaces and a symmetric distance derived from that correspondence.

Correspondence-based methods in shape analysis instead construct mappings between objects, ranging from rigid and non-rigid registration to functional maps and related spectral formulations~\cite{fischler1981random,besl1992method,chen1992object,amberg2007optimal,myronenko2010point,ovsjanikov2012functional,van2011survey,deng2022survey}. Protein surfaces, however, are rough, often non-isometric, and may contain cavities and narrow channels, making stable geometric landmarks difficult to define. More importantly, interaction specificity depends on chemical patterning as well as shape~\cite{mcbride2022}. The missing object is therefore a correspondence between complete protein surfaces determined jointly by intrinsic geometry and distributed physicochemical fields, together with a symmetric distance derived from that correspondence.

Conceptually, a protein surface is a curved manifold with intrinsic geometric distances and spatially organized physicochemical fields. A meaningful notion of similarity therefore relates fields across the underlying geometry. This requires a correspondence between surfaces: without one, geometric distortion and physicochemical mismatch cannot be combined into a single distance.

Here we introduce IFACE (Intrinsic Field--Aligned Coupled Embedding), which realizes this construction through a probabilistic coupling that provides bidirectional soft correspondences and defines a symmetric geometric--chemical distance between complete protein surfaces.

\begin{figure*}[htbp!]
    \centering
        \includegraphics[
            width=\textwidth,
            clip
        ]{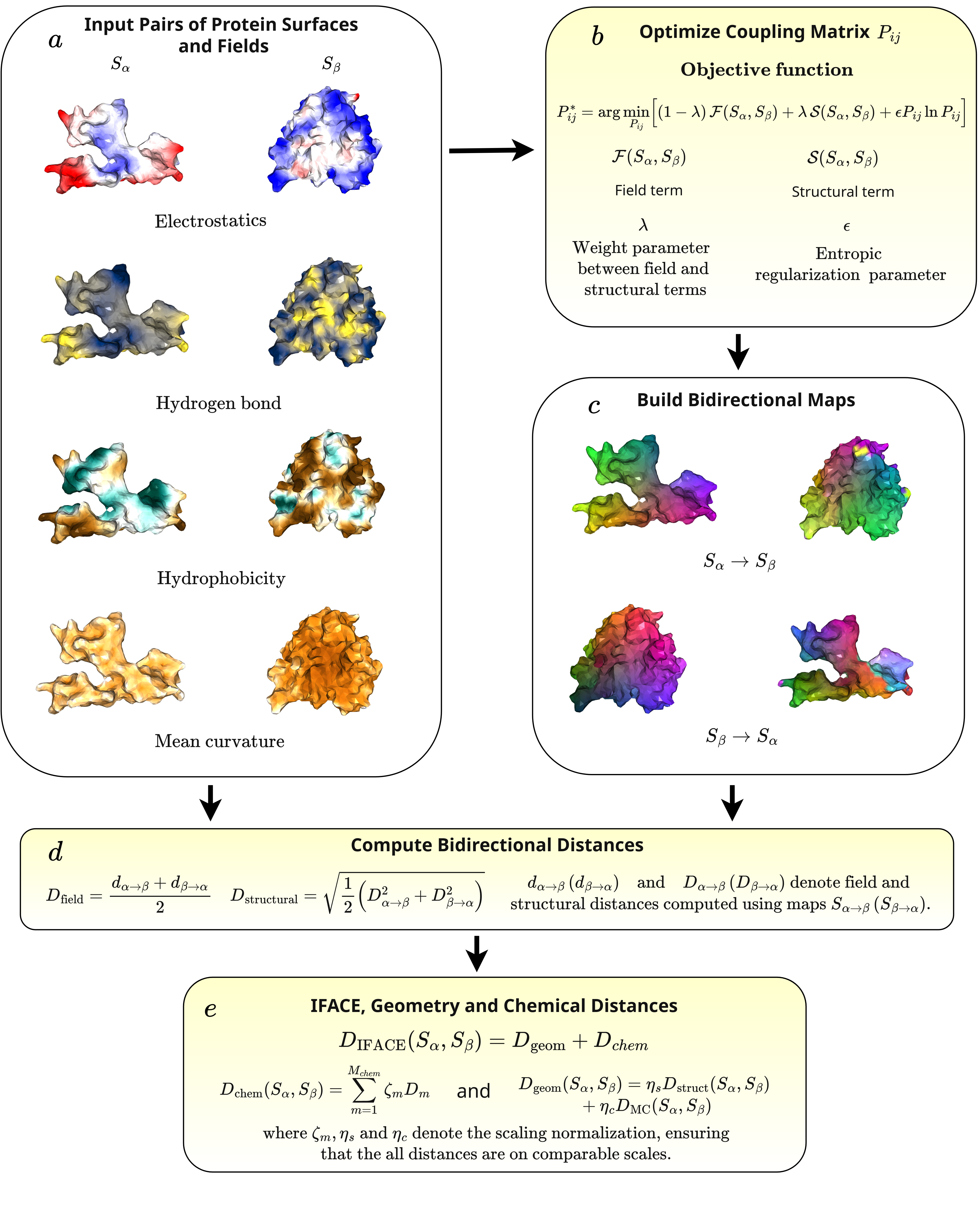}
 \caption{
\textbf{Conceptual workflow of IFACE.}
(a) Protein surfaces $S_{\alpha}$ and $S_{\beta}$ are represented using
geometry and surface feature fields (electrostatics, hydrogen-bond propensity,
hydrophobicity, and curvature).
(b) An optimal coupling matrix $P_{ij}$ is computed by balancing structural
and feature-field similarity.
(c) The coupling defines bidirectional soft correspondences between surfaces.
(d) Correspondence-based structural and feature-field distances are computed. (e) Distances are normalized to ensure they are all on the same scale; their combination defines the IFACE distance, which comprises a geometric distance (structural and mean-curvature [MC] components) and a chemical distance (hydrogen-bond, hydrophobicity, and electrostatic components).
}~\label{fig:visual_pipeline}
\end{figure*}

We demonstrate the framework in two complementary settings. First, we test whether the method distinguishes conformers of the same protein from different proteins. In this benchmark, IFACE combines intrinsic geometric distortion with mismatches in mean curvature, hydrogen-bond propensity, hydrophobicity, and electrostatic potential. It separates same-protein conformers from distinct proteins more accurately than TM-distance ($1-\mathrm{TM\text{-}score}$)~\cite{zhang2005tm} and a Laplace--Beltrami spectral distance. A correspondence-free Jensen--Shannon distribution distance performs best in this binary identity test, showing that aggregate surface-feature distributions can already encode protein identity.

Second, we examine similarity at the family level. Here the task is not merely to distinguish same from different: the full set of pairwise distances must organize many distinct protein surfaces according to their mutual relations. Across six protein families, IFACE gives the strongest classification and clustering among the Jensen--Shannon distribution, Laplace--Beltrami spectral, MaSIF, and SurfaceID distances examined. The benchmark includes cytochrome P450 proteins, whose shared fold encompasses substantial functional and surface diversity~\cite{guengerich2008cytochrome,omura2010structural}.

Finally, we evaluate the correspondence itself. IFACE preserves local geodesic neighborhoods and transfers heme-centered pocket regions between cytochrome P450 surfaces. Thus, the same variational construction provides both a geometric--chemical distance between complete protein surfaces and an explicit local map that reveals the surface regions underlying that distance.\\

\noindent\sb{The IFACE Method.} We introduce IFACE (Intrinsic Field--Aligned Coupled Embedding), a correspondence-based framework for comparing protein surfaces via coupled geometry and chemical fields. Figure~\ref{fig:visual_pipeline} summarizes the construction. Given two protein surfaces, $S_{\alpha}$ and $S_{\beta}$, each is represented by its intrinsic geometric structure together with spatially distributed surface fields: mean curvature, electrostatic potential, hydrogen-bond propensity, and hydrophobicity (Fig.~\ref{fig:visual_pipeline}a). Mean curvature is treated as a field because it is a local scalar measure of extrinsic geometry, distinct from the intrinsic geodesic structure of the surface. We then compute an optimal coupling matrix that aligns the two surfaces by balancing structural consistency with feature-field agreement (Fig.~\ref{fig:visual_pipeline}b). The relative contribution of these terms is controlled by the parameter $\lambda$, defined in the variational formulation below.

The resulting coupling defines soft correspondences between surface points and induces bidirectional surface maps between the two proteins (Fig.~\ref{fig:visual_pipeline}c). Using these correspondences, we compute symmetric structural and feature-field discrepancies (Fig.~\ref{fig:visual_pipeline}d). From these quantities, we construct a geometric distance combining intrinsic structural distortion with mean-curvature mismatch, and a chemical distance combining hydrogen-bond propensity, hydrophobicity, and electrostatic mismatch. After normalization to a common scale, their combination defines the geometric--chemical IFACE distance. This procedure yields a symmetric measure that integrates intrinsic and extrinsic geometry with aligned chemical fields while retaining the explicit surface correspondence. Full details of the optimization and definitions are provided in the Methods.

 Our implementation depends on a few parameter choices, most notably $\lambda$ and $\epsilon$ (Fig.~\ref{fig:visual_pipeline}b). The parameter $\lambda$ controls the balance between intrinsic geometry and surface-field agreement, whereas $\epsilon$ controls the strength of entropic regularization during optimization. Across broad ranges of both parameters, the principal results and relative performance of the distances remain stable (see SI).
\\

\section{\sb{Results}}

We evaluate IFACE in three stages: discrimination of molecular-dynamics conformers, organization of proteins across six families, and validation of the inferred surface correspondences. Unless stated otherwise, all surfaces are represented by meshes with \num{3000} vertices; details of surface generation are provided in the Supplementary Information. All distances are computed directly from the variational correspondence, without training or learned embeddings.

We begin with a basic but essential consistency test. If the representation faithfully encodes surface organization, distinct conformers of the same protein should remain close to one another, while remaining clearly separated from unrelated proteins, under the IFACE distance. We therefore compare IFACE distances among conformers of the same protein with distances between different proteins. This experiment probes whether conformational variability is correctly distinguished from genuine inter-protein dissimilarity.

We then consider a more demanding regime: organization at the protein-family level across six families. Here, we test whether proteins belonging to the same family group together under the IFACE distance. This setting asks whether the complete matrix of pairwise distances recovers shared family-level surface organization across diverse proteins. Successful classification and clustering indicate that the representation captures conserved geometric--chemical patterns rather than relying on any single global descriptor.

Finally, we evaluate the inferred surface correspondences directly. We test whether mapped vertices preserve local geodesic neighborhoods across the conformer and family benchmarks, and whether the maps transfer heme-centered pocket regions between cytochrome P450 proteins. These analyses assess whether the correspondence underlying the IFACE distance is locally coherent and identifies biologically corresponding surface regions.\\

\nsb{IFACE discerns conformers of the same protein as closer than different proteins.}
In this task, we use the ATLAS dataset~\cite{vander2024atlas} and focus on four proteins with
PDB codes~\cite{berman2000protein} \texttt{6XRX}~\cite{fooMosquito2021},
\texttt{5HZ7}~\cite{berryComparative2016},
\texttt{2XZ3}~\cite{lambChargeSurrounded2011},
and \texttt{6XDS}~\cite{byrneDevelopment2021},
corresponding to an AEG12 mosquito protein, a DNA-binding protein,
a viral protein, and a signaling protein, respectively. 
These structures are fusion constructs in which the proteins of interest are fused to maltose-binding protein (MBP), a commonly used solubility and crystallization tag. Consequently, each contains the same MBP domain fused to a different protein of interest.
All structures were restricted to chain~A.
ATLAS contains molecular dynamics (MD) trajectories for \num{1390} proteins from the Protein Data Bank (PDB), with three independent simulations per protein and 10,001 snapshots per trajectory. Each trajectory spans \SI{100}{ns}. On this timescale, the native fold is preserved; what fluctuates is the surface. 
The resulting ensembles therefore sample thermal surface variability without undergoing large conformational rearrangements. 
This makes ATLAS suitable for assessing whether a distance captures intrinsic surface fluctuations rather than gross structural change.

To construct a stringent benchmark, we searched for proteins that allow controlled comparison both within and between proteins. Sequence analysis using MMseqs2~\cite{steinegger2017mmseqs2} showed that most ATLAS entries are mutually unrelated, aside from occasional duplicates. A notable exception is a set of four related synthetic constructs fused to maltose-binding protein for crystallization: \texttt{6XRX\_A}~\cite{fooMosquito2021}, \texttt{5HZ7\_A}~\cite{berryComparative2016}, \texttt{2XZ3\_A}~\cite{lambChargeSurrounded2011}, and \texttt{6XDS\_A}~\cite{byrneDevelopment2021}.
These proteins are similar enough to make discrimination nontrivial, yet distinct enough to test whether combining surface chemistry with geometry adds resolving power beyond global fold similarity.

For each protein, we select ten conformers sampled along the MD trajectory. The resulting TM-score distributions overlap both within and between proteins, creating an ensemble in which intra-protein variability competes directly with inter-protein similarity. This overlap makes the benchmark demanding: it asks whether a surface-based optimal coupling, sensitive to both geometry and physicochemical fields, can resolve distinctions that fold-level similarity cannot.\\

\begin{figure*}[htbp!]
    \centering
     \includegraphics[width=0.9\textwidth]{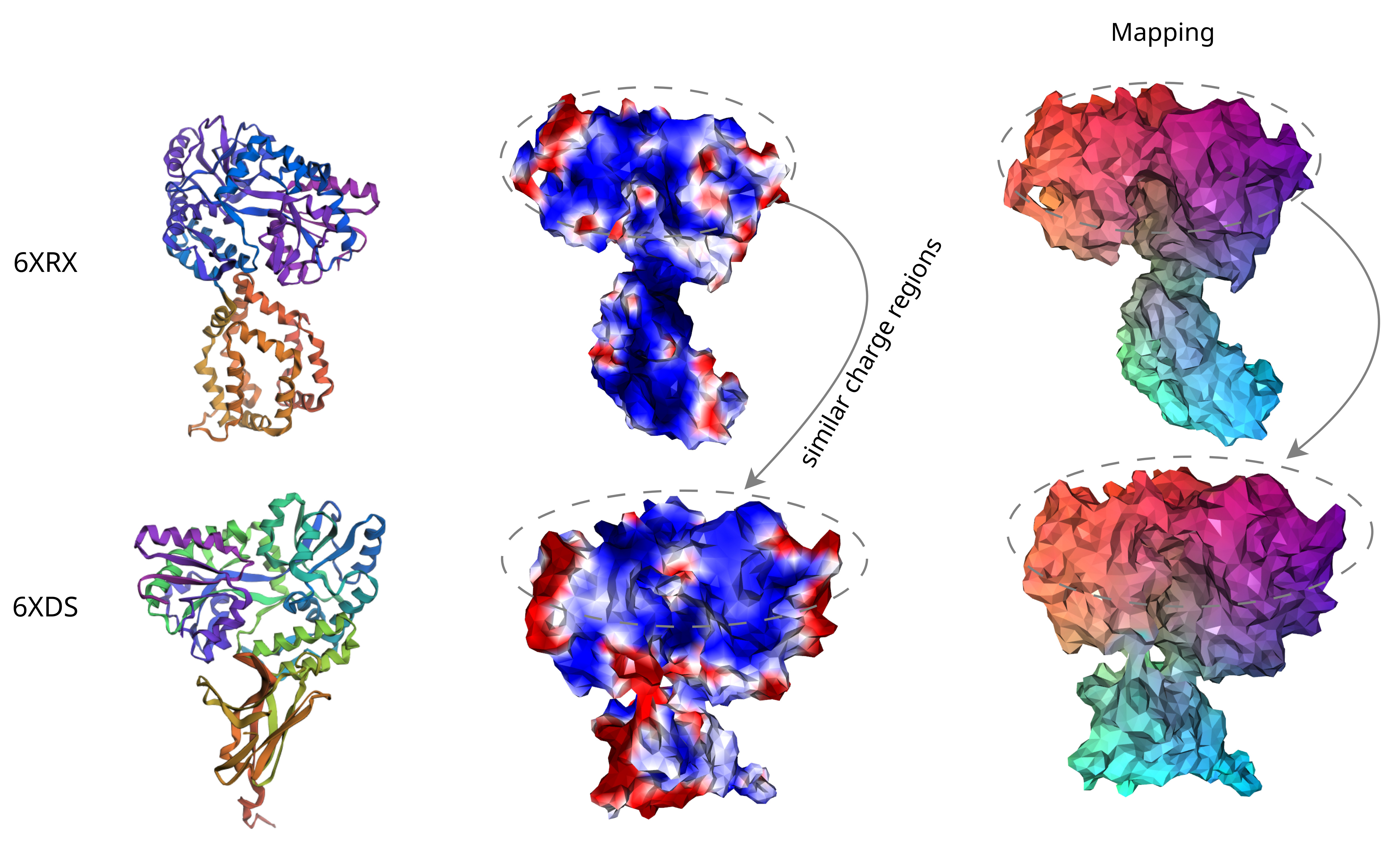}

\caption{
\textbf{Comparison and mapping of protein surfaces.}
Protein \texttt{6XRX} is shown in the top row and \texttt{6XDS} in the bottom row.
Left: ribbon representations of the protein structures.
Middle: electrostatic potential distributions (color scale clipped to the
5th–95th percentile of the combined distribution), where similar colors
indicate regions with comparable electrostatic potential.
Right: IFACE color mappings showing correspondences from \texttt{6XRX} to \texttt{6XDS}. The correspondences (vertex mapping) are computed using combined structural and
feature-field similarity, allowing direct comparison of corresponding
surface regions. 
}
  \label{fig:combined-protein}
\end{figure*}

\nsb{Example of IFACE mapping.}
Before turning to quantitative analysis, we first examine a representative example to visualize the surface correspondence identified by the IFACE optimal transport coupling (Fig.~\ref{fig:combined-protein}). The mapping integrates geometric alignment with multiple surface fields---hydrophobicity, hydrogen-bond propensity, electrostatic potential, and mean curvature---as described above.

The two proteins shown (\texttt{6XRX} and \texttt{6XDS}) have a TM-distance of \num{\sim 0.30} (TM-score $\sim 0.70$), indicating substantial global fold similarity. This similarity arises because both constructs contain the same fused maltose-binding protein (MBP) domain, which is identical in sequence in the two structures. 
The shared domain accounts for a substantial fraction of each construct and therefore produces strong global alignment. The non-MBP proteins have distinct biological functions, so the informative differences lie primarily outside the shared domain, in the non-MBP regions and their exposed surfaces.

The middle column of Fig.~\ref{fig:combined-protein} displays the electrostatic potential mapped onto the two surfaces. The upper regions of \texttt{6XRX} (lipid-binding protein) and \texttt{6XDS} (signaling protein) exhibit closely matched electrostatic profiles and compatible geometry. The IFACE correspondence (rightmost panel) aligns these regions across the two proteins, identifying surface patches that are jointly similar in shape and physicochemical character. In contrast, the lower regions differ in both curvature and electrostatics, and consequently contribute more strongly to the overall distance.

This example illustrates how the optimized surface correspondence integrates geometric and chemical information to resolve surface-level similarity beyond fold alignment alone. Having established this qualitative intuition, we now proceed to a systematic evaluation of whether IFACE reliably separates conformers of the same protein from distinct proteins.\\

\begin{figure*}[t]
\centering

{\fontsize{10}{12}\selectfont\textbf{Conformer-Benchmark Classification and Ablation}}\\[3mm]

\begin{minipage}{0.42\textwidth}
    \centering
    \includegraphics[width=\linewidth]{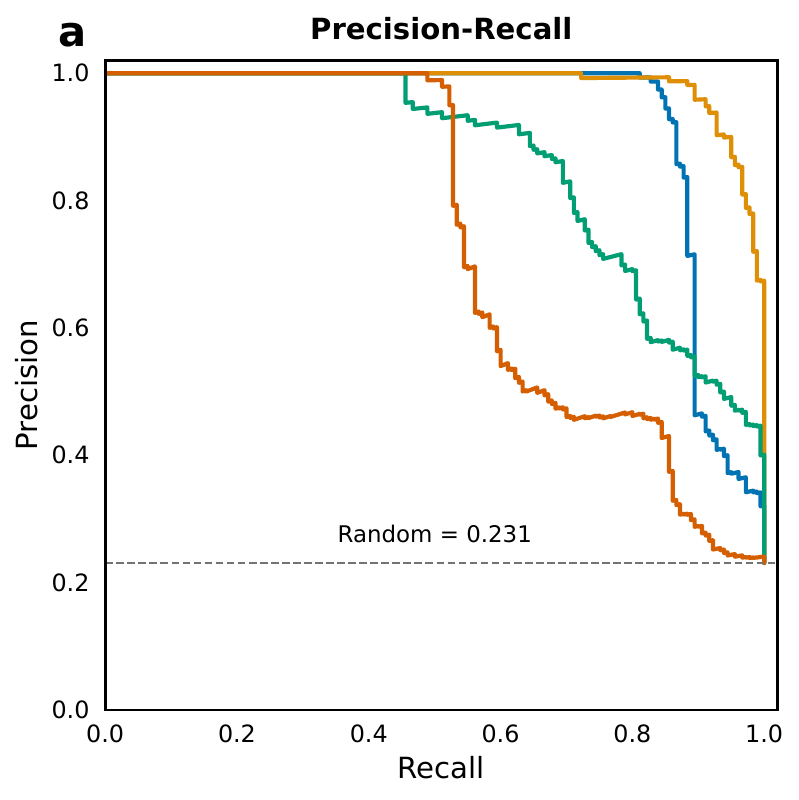}
\end{minipage}
\hfill
\begin{minipage}{0.14\textwidth}
    \centering
    \includegraphics[width=\linewidth]{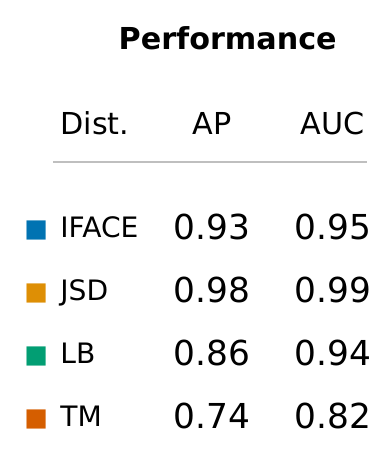}
\end{minipage}
\hfill
\begin{minipage}{0.42\textwidth}
    \centering
    \includegraphics[width=\linewidth]{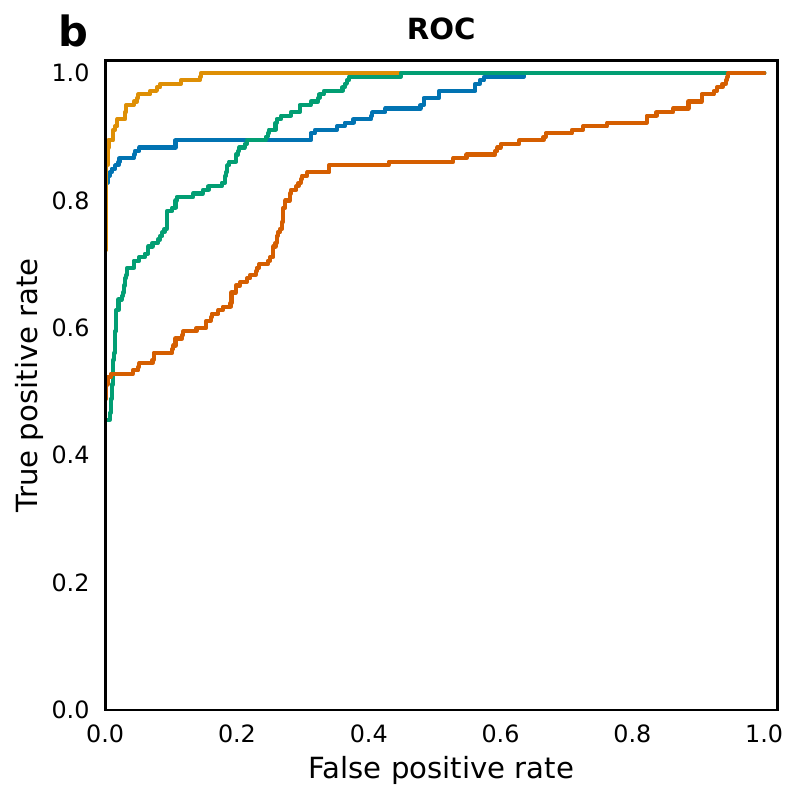}
\end{minipage}

\vspace{2mm}

\centering
\includegraphics[width=\linewidth]{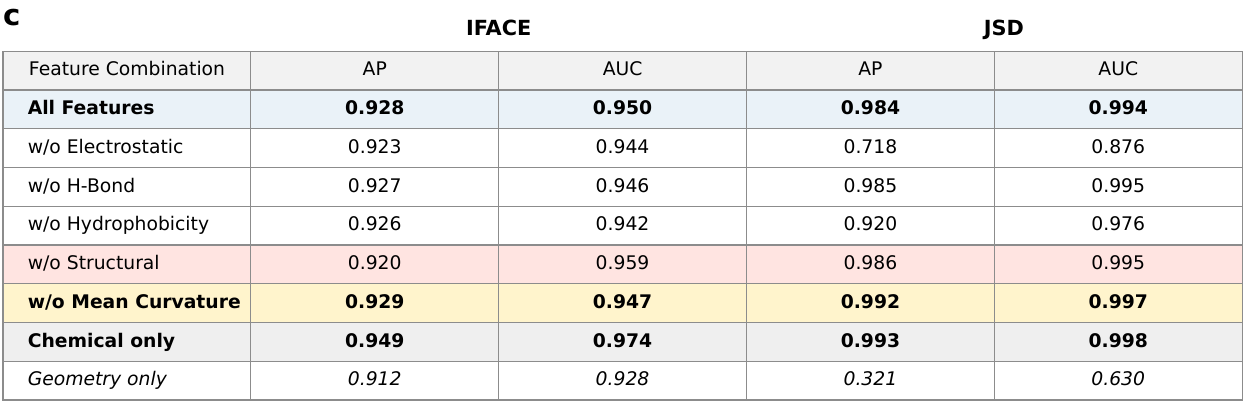}
\caption{
\textbf{Conformer benchmark classification and feature ablation.}
(a) Precision--Recall curves and (b) Receiver--Operating Characteristic (ROC)
curves comparing IFACE, JSD, Laplace--Beltrami (LB), and TM distances.
Performance is summarized by Average Precision (AP) and ROC--AUC. IFACE
outperforms both TM and LB, while JSD achieves the highest
classification performance on this benchmark. (c) Leave-one-feature-out
ablation comparing correspondence-based IFACE with matched distribution-based
JSD distances. Removing individual descriptors has only a modest effect on
IFACE, indicating that its performance is robust to individual feature
removal. In contrast, JSD exhibits greater sensitivity to descriptor
selection: removing the electrostatic feature substantially reduces
performance, while the geometry-only representation performs markedly worse
than its correspondence-based IFACE counterpart.
}
\label{fig:conformer_ablation}
\end{figure*}

\nsb{IFACE distance robustly separates conformers from distinct proteins.}
We quantify discriminatory power using the \emph{IFACE distance}, a composite measure that integrates geometric and physicochemical features of protein surfaces. Each component—structural (\texttt{struct}), mean curvature (\texttt{meancurv}), hydrophobicity (\texttt{hphob}), hydrogen-bond propensity (\texttt{hbond}), and electrostatic potential (\texttt{elect})—is normalized using bounds obtained from a fixed reference set, placing the components on comparable numerical scales.

The IFACE distance is then defined as 
\begin{equation}
\label{eq:iface}
   \text{iface}
~=~
\text{struct}+\text{meancurv}+\text{hphob}+\text{hbond}+\text{elect}~. 
\end{equation}
This assigns equal weight to intrinsic structure, mean curvature, and each physicochemical field.
This equal-weight choice avoids parameter tuning and provides a transparent baseline against which more elaborate weighting schemes could be assessed (Eq.~\ref{eq:iface} corresponds to Eq.~\ref{eq:iface_distance} in Methods with equal weights).
For comparison, we also define a \emph{chemical-feature distance},
\begin{equation*}
\Label{eq:chemical}
    \text{chemical}
~=~
\text{hphob}+\text{hbond}+\text{elect}
,
\end{equation*}
which isolates the contribution of physicochemical fields independent of structural alignment.
We also define a \emph{geometry-only distance},
\begin{equation*}
\Label{eq:geometry}
    \text{geom}
~=~
\text{struct}+\text{meancurv}~,
\end{equation*}
which isolates the contribution of geometry from that of the chemical fields.

For this task, we compare IFACE against three classes of baselines: 
(i) TM-distance, defined as $1-\mathrm{TM\mbox{-}score}$, where
TM-score is a widely used structural similarity measure based on structural alignment; 
(ii) a correspondence-free distributional baseline defined as the average Jensen--Shannon divergence~\cite{lin2002divergence} computed independently for each geometric and
physicochemical surface feature; and 
(iii) an intrinsic spectral distance based on the Laplace--Beltrami (LB) operator~\cite{reuter2006laplace}. These baselines provide complementary comparisons based on global
structural similarity, feature distributions, and intrinsic geometry,
respectively.

Figure~\ref{fig:conformer_ablation} summarizes the binary classification performance and feature ablation analysis on the conformer benchmark. 
For the componentwise ablation comparison, the four scalar surface fields are matched directly between IFACE and JSD. Because the intrinsic structural component of IFACE has no direct distributional counterpart, it is compared with the JSD of the Gaussian-curvature distribution as an additional geometric descriptor.
The precision--recall (Fig.~\ref{fig:conformer_ablation}a) and ROC (Fig.~\ref{fig:conformer_ablation}b) curves show that both IFACE and the distribution-based JSD representation accurately distinguish
same-protein conformers from distinct proteins,
 outperforming the Laplace--Beltrami (LB) distance and the TM-score baseline. 
For this benchmark, JSD achieves the highest overall performance (AP = 0.984 vs.~0.928; ROC-AUC = 0.994 vs.~0.950), indicating that aggregate descriptor distributions are especially effective in this binary protein-identity test.

The feature contribution analysis (Fig.~\ref{fig:conformer_ablation}c) provides further insight into the role of individual descriptors.
Electrostatics is the dominant descriptor for JSD: removing it reduces AP from 0.984 to 0.718 and ROC-AUC from 0.994 to 0.876. The corresponding reduction for IFACE is only modest, from AP = 0.928 to 0.923 and ROC-AUC = 0.950 to 0.944. Removing the remaining descriptors likewise produces only minor variations in IFACE performance, whereas JSD varies more strongly for Electrostatics.

Chemical descriptors alone slightly outperform the full five-feature representation for both methods (IFACE: AP = 0.949, ROC-AUC = 0.974; JSD: AP = 0.993, ROC-AUC = 0.998). By contrast, geometry-only JSD performs substantially worse (AP = 0.321, ROC-AUC = 0.630), whereas geometry-only IFACE shows only a small degradation. 
Thus, JSD discrimination in this benchmark depends strongly on physicochemical information, particularly electrostatics, whereas no single descriptor is solely responsible for the performance of IFACE.
\\

\begin{table*}[ht]
\caption{\label{tab:benchmark_proteins}
Proteins used in the family-level clustering benchmark, their source
organisms, and common organism classifications.}
\begin{ruledtabular}
\begin{tabular}{llp{0.32\textwidth}p{0.18\textwidth}}
PDB ID & Protein family/group & Source organism & Common name/type \\
\hline

2IJ5 & Cytochrome P450
& \textit{Mycobacterium tuberculosis}
& Bacterium \\

6N6Q & Cytochrome P450
& Mycobacterium phage Adler
& Bacteriophage \\

3OFU & Cytochrome P450
& \textit{Novosphingobium aromaticivorans}
& Bacterium \\

6ZI2 & Cytochrome P450
& \textit{Streptomyces antibioticus}
& Bacterium \\

6UX0 & Cytochrome P450
& \textit{Acanthamoeba castellanii}
& Amoeba \\

4H23 & Cytochrome P450
& \textit{Priestia megaterium}
& Bacterium \\

1JPZ & Cytochrome P450
& \textit{Priestia megaterium}
& Bacterium \\

1TQN & Cytochrome P450
& \textit{Homo sapiens}
& Human \\
\hline

1AOI & Histone proteins
& \textit{Xenopus laevis}
& African clawed frog \\

5Y0C & Histone proteins
& \textit{Homo sapiens}
& Human \\

4QLC & Histone proteins
& \textit{Drosophila melanogaster};
  \textit{Gallus gallus}
& Fruit fly; chicken \\
\hline

1BTP & Serine proteases
& \textit{Bos taurus}
& Cattle \\

1ESA & Serine proteases
& \textit{Sus scrofa}
& Pig \\

1GCD & Serine proteases
& \textit{Bos taurus}
& Cattle \\

\hline

2DN1 & Globins
& \textit{Homo sapiens}
& Human \\

1A6M & Globins
& \textit{Physeter macrocephalus}
& Sperm whale \\

1OJ6 & Globins
& \textit{Homo sapiens}
& Human \\

1UT0 & Globins
& \textit{Homo sapiens}
& Human \\
\hline

1LYZ & C-type lysozymes
& \textit{Gallus gallus}
& Chicken \\

1LZ1 & C-type lysozymes
& \textit{Homo sapiens}
& Human \\


\hline

7RSA & Ribonuclease A superfamily
& \textit{Bos taurus}
& Cattle \\

1ANG & Ribonuclease A superfamily
& \textit{Homo sapiens}
& Human \\

1GQV & Ribonuclease A superfamily
& \textit{Homo sapiens}
& Human \\

2RNF & Ribonuclease A superfamily
& \textit{Homo sapiens}
& Human \\
\hline

3L40 & Outlier protein
& \textit{Schizosaccharomyces pombe}
& Fission yeast \\

\end{tabular}
\end{ruledtabular}
\end{table*}

\noindent\sb{Family-level similarity of proteins.}
In this experiment, we evaluate whether the proposed distance measures organize proteins coherently at the family level. 
Unlike the binary conformer benchmark, this task tests whether the full matrix of pairwise distances recovers relationships among many distinct protein surfaces.
This framework yields a distance-based representation of protein space and enables direct comparisons across protein families without relying on sequence or fold annotations.

To evaluate family-level protein surface similarity, we constructed a benchmark comprising 
25 proteins in total: 24 from six structurally and functionally distinct protein families and 
one unrelated outlier (PDB ID: 3L40)
(Table~\ref{tab:benchmark_proteins}). The benchmark includes eight cytochrome P450 proteins, three histone proteins, three serine proteases, four globins, two C-type lysozymes, and four members of the ribonuclease A superfamily, as well as a single unrelated outlier. These families were selected because they exhibit substantial diversity in biological function, sequence, and structural organization, while remaining sufficiently well characterized to provide representative examples of each family. 
The proteins originate from evolutionarily diverse organisms, including bacteria, a bacteriophage, an amoeba, humans, cattle, pigs, sperm whales, chickens, frogs, fruit flies, and fission yeast. This selection introduces considerable evolutionary diversity both within and across protein families.

Our benchmark is designed to determine whether surface-based similarity measures recover biologically meaningful family relationships rather than merely reflecting sequence identity or global structural resemblance. Proteins belonging to the same family are expected to cluster together despite evolutionary divergence, whereas proteins from different families should remain well separated. The inclusion of a structurally unrelated outlier further assesses the robustness of the distance measures when applied to proteins that do not naturally belong to any benchmark family. 
Together, these features make the set a stringent benchmark for family-level classification, clustering, and correspondence-based surface comparison.

\begin{figure*}[htbp!]
    \centering
    \includegraphics[width=0.8\textwidth]{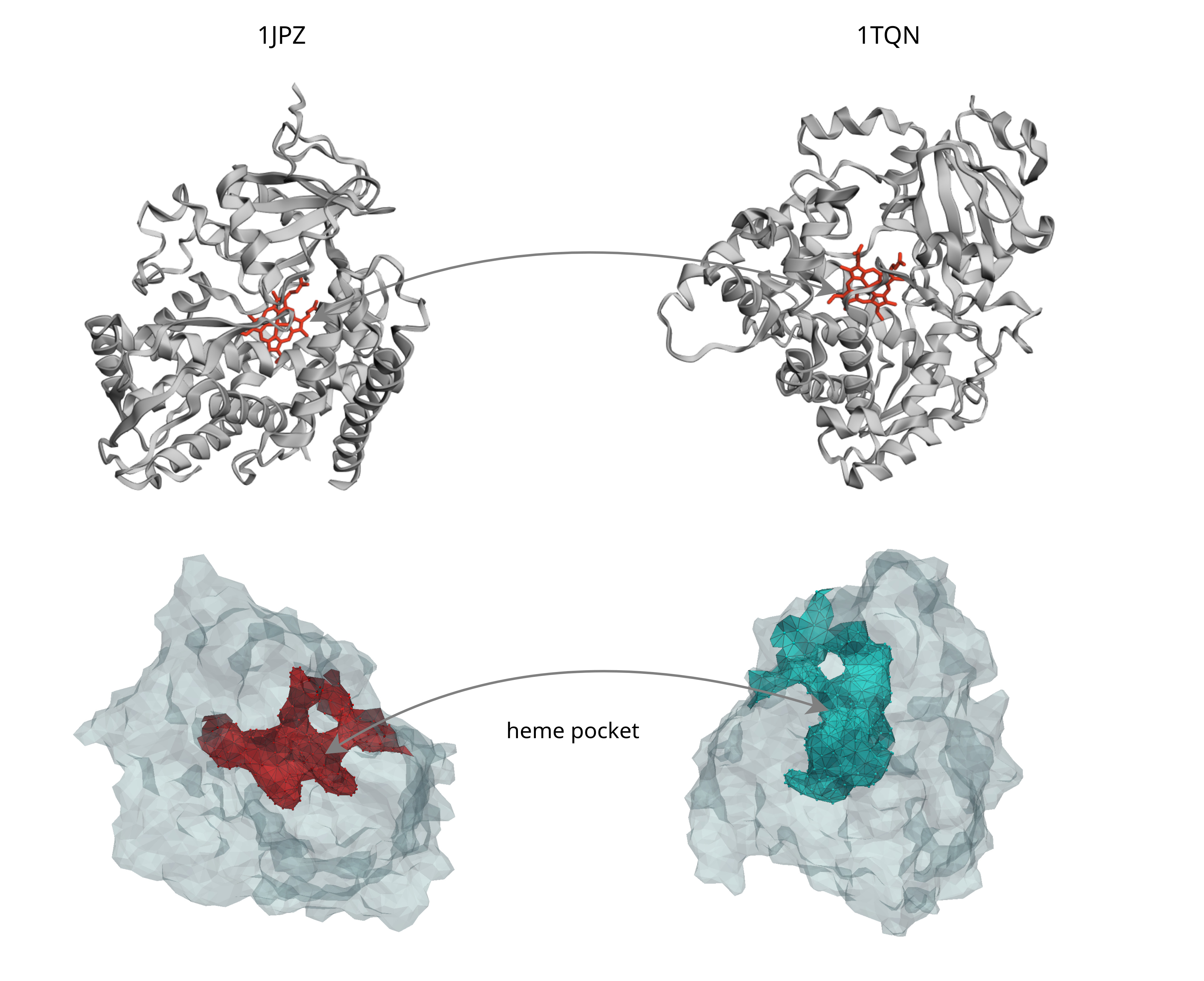}
  \caption{
\textbf{Comparison of proteins \texttt{1JPZ} and \texttt{1TQN}.}
Top row: ribbon representations of chain A from \texttt{1JPZ} (left)
and \texttt{1TQN} (right), with the heme prosthetic group shown in red.
Bottom left: patch around the heme on \texttt{1JPZ} (red), containing the
heme group responsible for substrate oxidation via oxygen activation.
This patch lies within the protein interior and is accessed through
substrate tunnels, illustrating non-trivial internal surface geometry.
Bottom right: the region on 1TQN obtained by mapping the heme-centered pocket of 1JPZ with IFACE, illustrating correspondence between the buried pocket regions of the two proteins.
}
  \label{fig:pocket protein-1tqn-1jpz}
\end{figure*}

\nsb{Example of pocket mapping within the same protein family.}
We illustrate surface correspondence within the P450 family 
using full-length human cytochrome P450 3A4 (PDB ID: \texttt{1TQN}, sequence length 486, chain A) and P450 BM-3 from \textit{Bacillus megaterium} (PDB ID: \texttt{1JPZ}, sequence length 473, chain A). Both proteins contain a heme prosthetic group coordinated by a conserved cysteine residue. The heme pocket is deeply buried, with substrate access occurring through channels or tunnels. Its internal topology—comprising cavities, handles, and narrow passages—renders it difficult to identify using surface-comparison schemes such as conformal mapping or projection onto a plane or sphere.

Figure~\ref{fig:pocket protein-1tqn-1jpz} compares proteins \texttt{1JPZ} and \texttt{1TQN}. The top row shows ribbon representations of \texttt{1JPZ} (left) and \texttt{1TQN} (right), with the heme prosthetic group highlighted in red. 
The bottom-left image highlights the surface patch surrounding the heme in \texttt{1JPZ} in red. This buried patch is accessed through substrate tunnels and surrounds the catalytic heme, where substrate oxidation occurs through oxygen activation, exemplifying nontrivial internal surface geometry.
The bottom-right image shows the corresponding mapped region in \texttt{1TQN} (blue), identifying the homologous buried pocket.

The probabilistic surface correspondence recovers this conserved internal organization despite substantial geometric complexity. The alignment identifies 
corresponding heme-centered regions 
without requiring global topological simplification, indicating that the coupling captures intrinsic surface organization rather than superficial shape similarity.
\\

\begin{figure*}[t]
\centering
{\fontsize{10}{12}\selectfont\textbf{Family-Benchmark Classification and Ablation}}\\[3mm]

\begin{minipage}{0.42\textwidth}
    \centering
    \includegraphics[width=\linewidth]{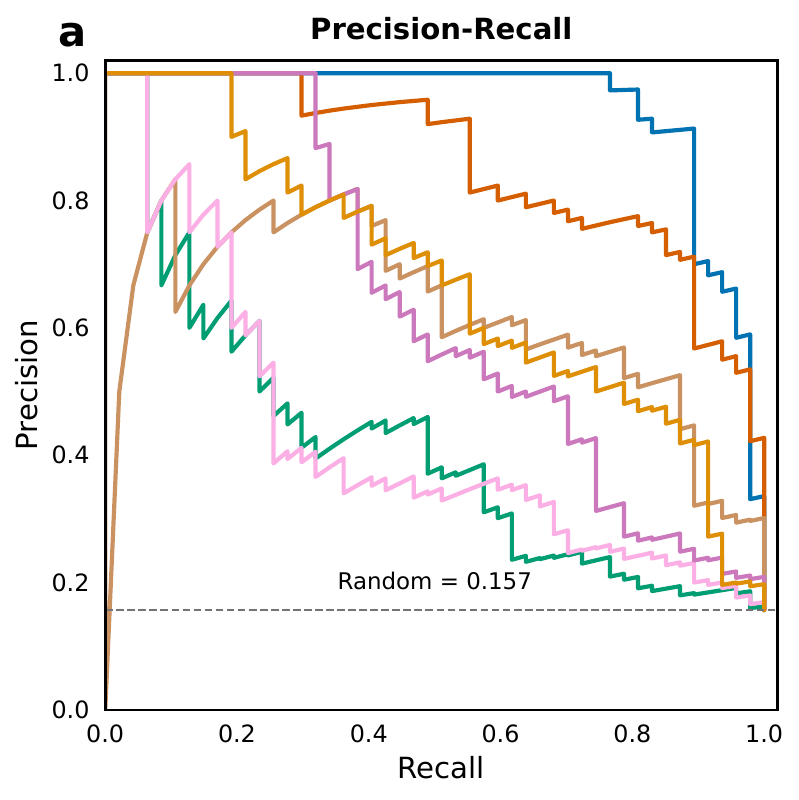}
\end{minipage}
\hfill
\begin{minipage}{0.14\textwidth}
    \centering
    \includegraphics[width=\linewidth]{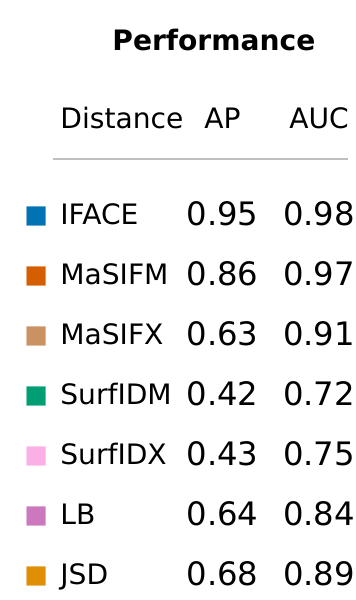}
\end{minipage}
\hfill
\begin{minipage}{0.42\textwidth}
    \centering
    \includegraphics[width=\linewidth]{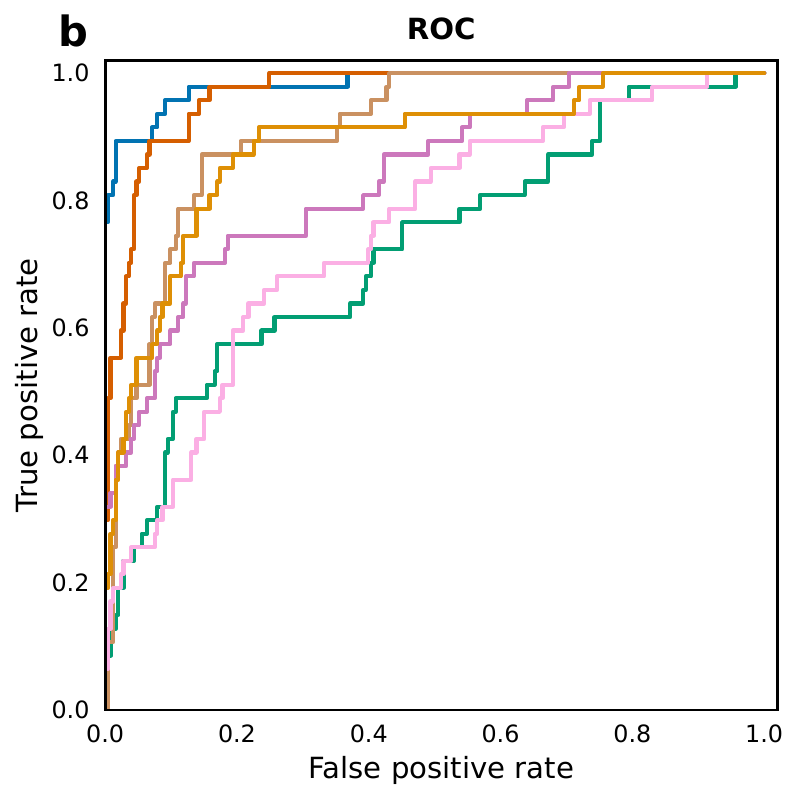}
\end{minipage}

\vspace{2mm}

\centering
\includegraphics[width=\linewidth]{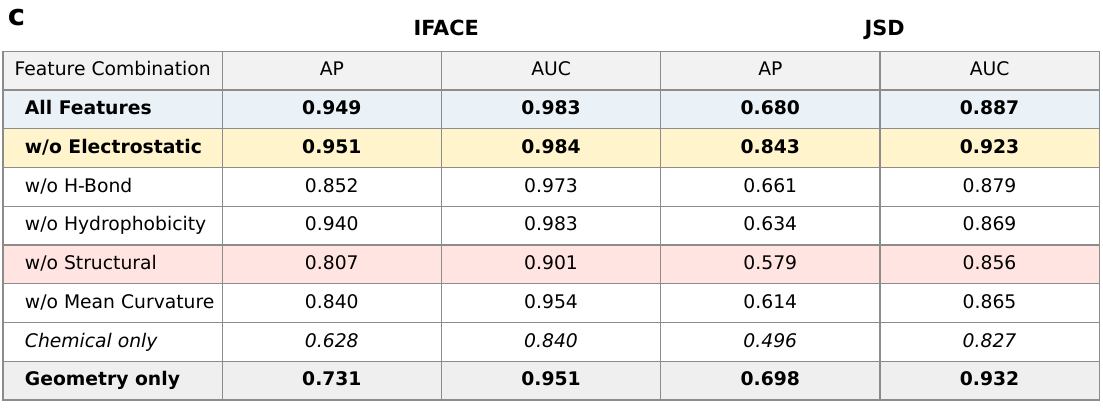}

\caption{
\textbf{Family benchmark classification and feature ablation.}
(a) Precision--Recall curves and 
(b) Receiver--Operating Characteristic (ROC) curves comparing IFACE with MaSIF (mean (M) and mean--max (X) pooling), SurfaceID (mean (M) and mean--max (X) pooling), Laplace--Beltrami (LB), and JSD distances for discriminating intra-family from inter-family protein pairs. Performance is summarized by Average Precision (AP) and ROC--AUC. IFACE achieves the best overall classification performance among all evaluated methods.  
(c) Leave-one-feature-out ablation comparing correspondence-based IFACE with matched JSD distances.
Removing the structural descriptor produces the largest overall decrease in IFACE performance, highlighting the dominant contribution of intrinsic geometry to family-level discrimination. Among the chemical descriptors, removing hydrogen-bond propensity causes the largest performance decrease, whereas removing electrostatics yields a marginally higher AP than the full feature set.
}
\label{fig:family_ablation}
\end{figure*}

\clearpage
\noindent\sb{Surface-distance-based clustering of protein surfaces.}
In this task, using the family benchmark, we compare protein pairs from the same family (intra-family) with pairs drawn from different families (inter-family). As shown in Fig.~\ref{fig:family_ablation}(a), 
we formulate the comparison as a binary classification task, with intra-family pairs as positives and inter-family pairs as negatives.

Beyond the distribution-based (JSD) and intrinsic spectral Laplace--Beltrami baselines, we additionally considered two state-of-the-art learning-based surface representation methods for the family-level benchmark. We used the released pretrained models of MaSIF-search~\cite{gainzaDeciphering2020} and SurfaceID~\cite{riahi2023surface} without additional training or fine-tuning to extract local descriptors from the protein surfaces. Since both methods were designed to learn local surface representations rather than whole-protein descriptors, their descriptors were aggregated to enable global protein comparison.

For the learning-based baselines, a fixed-length representation was constructed for each protein surface by aggregating the local patch descriptors produced by the pretrained models. 
MaSIF-search, trained for protein--protein interface prediction, produces an 80-dimensional descriptor for each surface patch, whereas SurfaceID, trained by self-supervised learning, produces a 512-dimensional descriptor. Because the number of patches varies among proteins, we evaluated two aggregation schemes. In the Mean variant, patch descriptors were averaged over the surface. In the Mean-Max variant, the element-wise mean and maximum were concatenated, yielding 160-dimensional MaSIF-search and 1024-dimensional SurfaceID representations. Pairwise protein distances were then defined as Euclidean distances between these global descriptors.\\

Figure~\ref{fig:family_ablation} summarizes the binary classification performance and feature ablation analysis on the family benchmark. 
The precision--recall (Fig.~\ref{fig:family_ablation}a) and ROC (Fig.~\ref{fig:family_ablation}b) curves demonstrate that IFACE achieves the highest overall performance (AP = 0.95, ROC-AUC = 0.98), outperforming both correspondence-independent baselines, JSD and Laplace--Beltrami (LB), and the learning-based surface descriptors MaSIF Mean (MaSIFM), MaSIF Mean-Max (MaSIFX), SurfaceID Mean (SurfIDM), and SurfaceID Mean-Max (SurfIDX).

Although these learned descriptors were originally developed for local patch comparison and descriptor retrieval rather than whole-protein family classification, they provide useful baselines for assessing how well pretrained local representations transfer to global protein comparison.
Unlike the conformer benchmark, where global feature distributions alone are sufficient to distinguish different protein identities, family-level discrimination benefits from preserving the spatial organization of geometric and physicochemical features across complete surfaces.

The feature contribution analysis (Fig.~\ref{fig:family_ablation}c) further supports this observation. In contrast to the conformer benchmark, removing the electrostatic descriptor substantially improves JSD performance, indicating that the marginal electrostatic distribution is comparatively poorly conserved within these protein families and may obscure family-level similarity.
Removing electrostatics from IFACE produces only a negligible improvement relative to the full distance.
Furthermore, removing either the structural or mean-curvature descriptor causes a larger decrease in IFACE performance than removing any individual chemical descriptor, indicating that geometry plays a more prominent role in family-level discrimination than in conformer recognition.
Interestingly, for IFACE, the geometry-only representation (AP = 0.731, ROC-AUC = 0.951) performs noticeably better than the chemical-only representation (AP = 0.628, ROC-AUC = 0.840).  

In contrast, combining these complementary sources of information through the full correspondence-aware IFACE representation increases the AP compared with using either modality alone. Although the full IFACE performance is only negligibly lower than that obtained without the electrostatic descriptor, the full representation
remains among the best-performing feature configurations, demonstrating that geometric and chemical information provide complementary contributions to family-level discrimination.

 A similar trend is observed for JSD, although both its geometry-only and chemical-only representations perform substantially worse than their IFACE counterparts. These observations indicate that explicit surface correspondences preserve complementary geometric and physicochemical information that is largely lost when descriptors are reduced to global feature distributions. Overall, the family benchmark demonstrates that correspondence-aware integration of geometry and chemistry becomes increasingly important as the comparison task shifts from distinguishing different protein identities to 
recognizing relationships among homologous proteins across diverse families.

We therefore ask whether the protein families emerge as coherent clusters in the full distance space. 
Table~\ref{tab:family_cluster_ari} summarizes both the peak clustering performance and the robustness of each distance measure across a diverse range of clustering strategies. To avoid bias toward a particular clustering algorithm, we evaluated four hierarchical linkage methods (single, complete,
average, and weighted) together with a broad collection of Classical Multidimensional Scaling (CMDS) pipelines. The CMDS evaluation included both fixed-dimensional embeddings (4, 5, 6, 8, 10, 15, and 20 dimensions) and variance-preserving embeddings retaining 80\%, 90\%, 95\%, and 99\% of the
explained variance, each followed by K-means and Ward clustering. To quantify algorithmic robustness, we report the coefficient of variation (CV = SD/Mean) computed separately for the hierarchical methods, the CMDS-based pipelines, and
all clustering settings combined, where a lower CV indicates lower sensitivity to the choice of clustering algorithm.

\begin{table*}[t]
\caption{
Summary of family-level clustering performance and robustness.
Best Hier. (hierarchical)\ and Best CMDS report the highest Adjusted Rand Index (ARI)
obtained using hierarchical and CMDS-based clustering, respectively.
Mean and coefficient of variation
(CV = SD/Mean) are reported separately for robustness for hierarchical clustering,
CMDS-based clustering, and all clustering settings. IFACE outperforms all other methods.
}
\centering
\includegraphics[width=\textwidth]{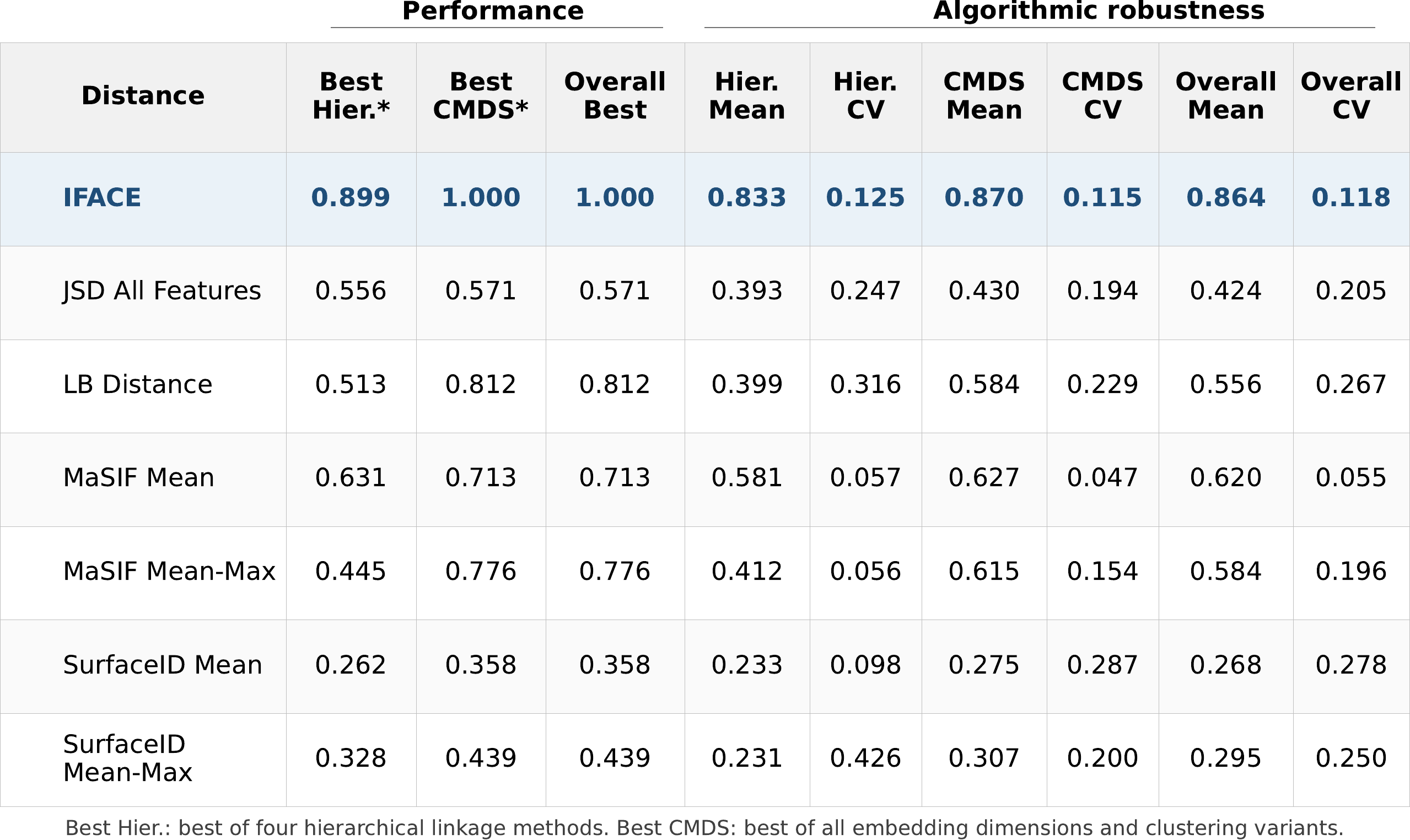}
\label{tab:family_cluster_ari}
\end{table*}

As shown in Table~\ref{tab:family_cluster_ari}, IFACE consistently achieves the strongest clustering performance among all evaluated methods. It attains the highest ARI under both hierarchical and CMDS-based clustering strategies, while also delivering the best overall mean ARI. In contrast, the distribution-based JSD distance exhibits substantially weaker clustering performance. Although the Laplace--Beltrami distance outperforms JSD, it remains consistently inferior to IFACE.

Among the learning-based baselines, MaSIF Mean achieves the strongest clustering performance, followed by MaSIF Mean--Max, whereas both SurfaceID variants perform considerably worse. Nevertheless, all learning-based baselines remain well below
IFACE. IFACE also demonstrates strong algorithmic robustness, maintaining consistently high clustering accuracy across different clustering algorithms with only limited sensitivity to algorithm selection. Although MaSIF Mean exhibits the lowest variability across clustering strategies, this stability is accompanied by consistently moderate clustering accuracy.
Overall, these results show that IFACE provides the best balance of clustering accuracy and algorithmic robustness across a diverse range of hierarchical and CMDS-based clustering approaches.

\begin{table*}[t]
\caption{
Summary of clustering performance and robustness of
IFACE distance ablation on the family benchmark.
Best Hier.\ (hierarchical) and Best CMDS report
the highest Adjusted Rand Index (ARI) obtained using hierarchical and
CMDS-based clustering, respectively. Mean and coefficient of variation
(CV = SD/Mean) summarize performance across hierarchical methods,
CMDS-based clustering, and all clustering settings. 
Removing the electrostatic descriptor modestly improves mean clustering performance, whereas removing the structural descriptor produces the largest degradation.}

\centering
\includegraphics[width=\textwidth]{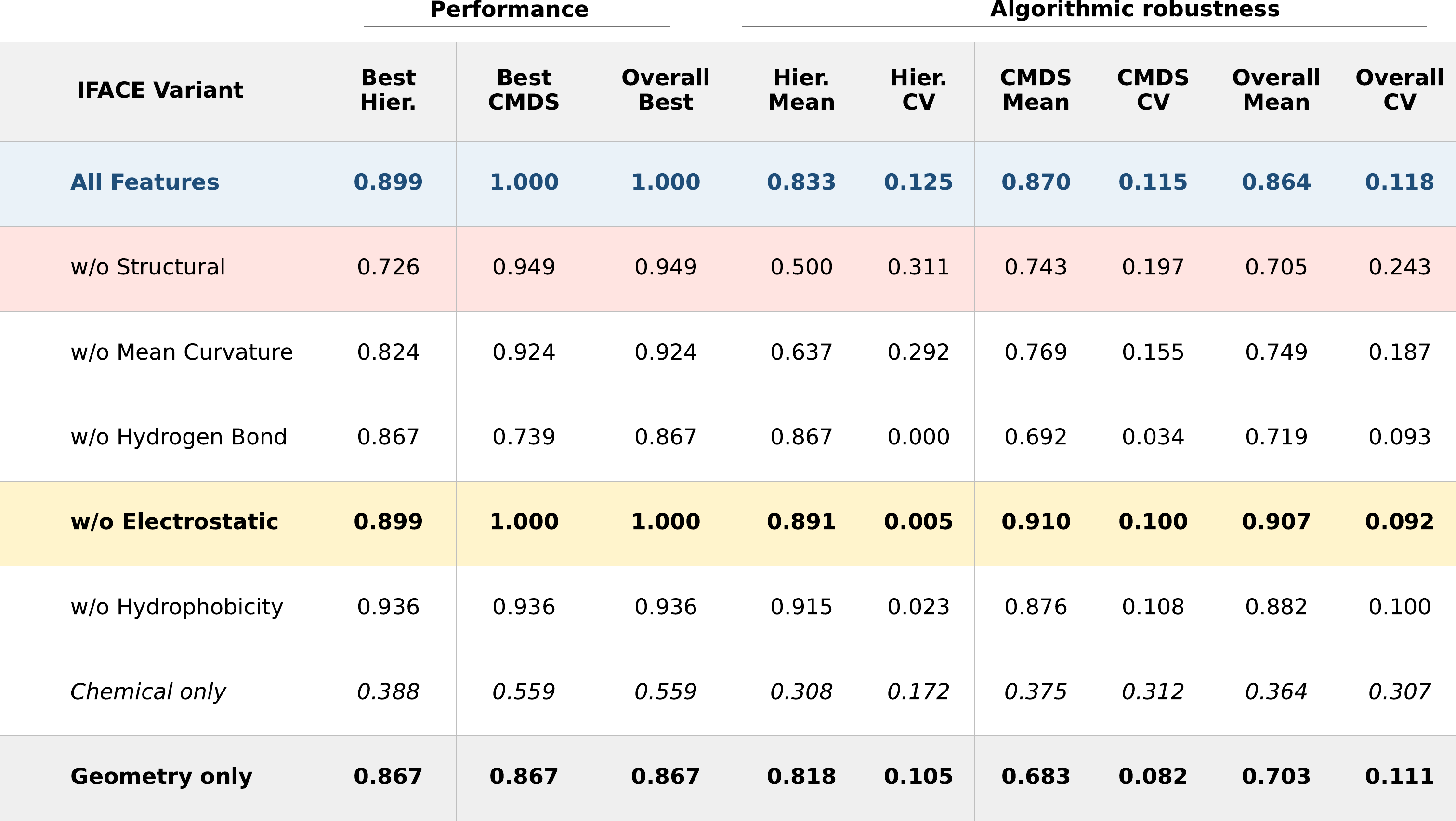}
\label{tab:family_cluster_ablation_h}
\end{table*}

Consistent with the results of the intra- versus inter-family classification task, the leave-one-feature-out ablation (Table~\ref{tab:family_cluster_ablation_h}) demonstrates that the contributions of the individual descriptors are not uniform. 
Removing the structural descriptor produces the largest degradation in clustering performance, reducing the overall mean ARI from 0.864 to 0.705, indicating that intrinsic surface geometry provides the largest individual contribution among the descriptors.

Nevertheless, geometry alone, comprising both intrinsic structural and extrinsic mean-curvature information, does not fully capture protein family similarity. Although the geometry-only variant substantially outperforms the chemical-only variant (overall mean ARI of 0.703  versus 0.364), it remains inferior to the complete IFACE distance (overall mean ARI of 0.864), demonstrating that geometric and physicochemical descriptors provide complementary information for protein family discrimination. 

Conversely, the chemical-only variant exhibits consistently weak clustering performance, indicating that physicochemical information alone is insufficient for reliable family-level discrimination within this benchmark.
However, removing the electrostatic descriptor modestly increases the overall mean ARI, whereas removing hydrophobicity produces a smaller improvement.

Overall, the complete IFACE distance performs strongly on both the conformer and family benchmarks, indicating that integrating intrinsic geometry, extrinsic geometry, and multiple physicochemical descriptors yields a robust protein-surface comparison distance without task-specific feature selection. These conclusions are further supported by the cluster-purity results (Supplementary Tables), where IFACE generally achieves the highest or among the highest purity scores across the evaluated clustering algorithms.

Furthermore, the corresponding leave-one-feature-out ablation for the distribution-based JSD distance (see Supplementary Information) is generally weaker than its IFACE counterpart, indicating that global feature distributions do not reproduce the same family-level organization. Complete leave-one-feature-out and feature-combination ablation studies are provided in the Supplementary Information.

Taken together, these results show that combining geometry with physicochemical surface fields enables IFACE to produce a stable and discriminative family-level organization within this benchmark.

\section{\sb{Effect of surface resolution}}

To assess whether the surface discretization was sufficient, we repeated the family- and conformer-level classification and clustering analyses using surfaces downsampled to 1,000 vertices and compared the results with those obtained using 3,000 vertices. Performance remained comparable across the two resolutions, indicating that the lower-resolution meshes preserve the geometric and physicochemical information required to distinguish protein families and conformational states. 

At 1,000 vertices, IFACE maintained performance comparable to that obtained at 3,000 vertices. The conformer benchmark yielded an AP of 0.91 and a ROC-AUC of 0.92, while the family benchmark yielded an AP of 0.97 and a ROC-AUC of 0.99. Hierarchical clustering achieved an ARI of 0.90, and the CMDS-based clustering results were similarly strong across several embedding dimensions. Thus, 1,000-vertex surfaces provide an adequate representation for these analyses while substantially reducing the computational cost of the correspondence calculation (see Section Scaling)
\section{\sb{P450 heme-pocket-active site transfer evaluation}}
\Label{sec:pocket_transfer}

\begin{figure*}[t]
\centering
\includegraphics[width=0.49\textwidth]{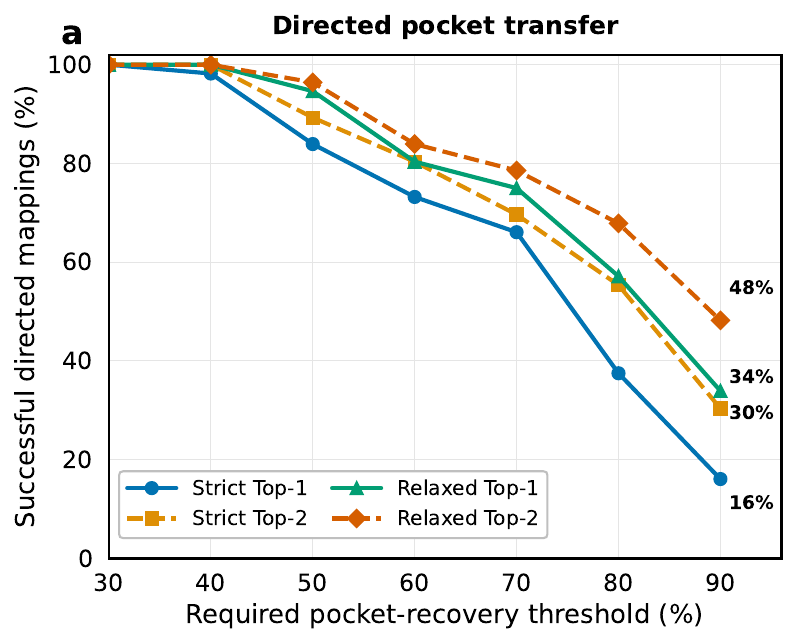}
\hfill
\includegraphics[width=0.49\textwidth]{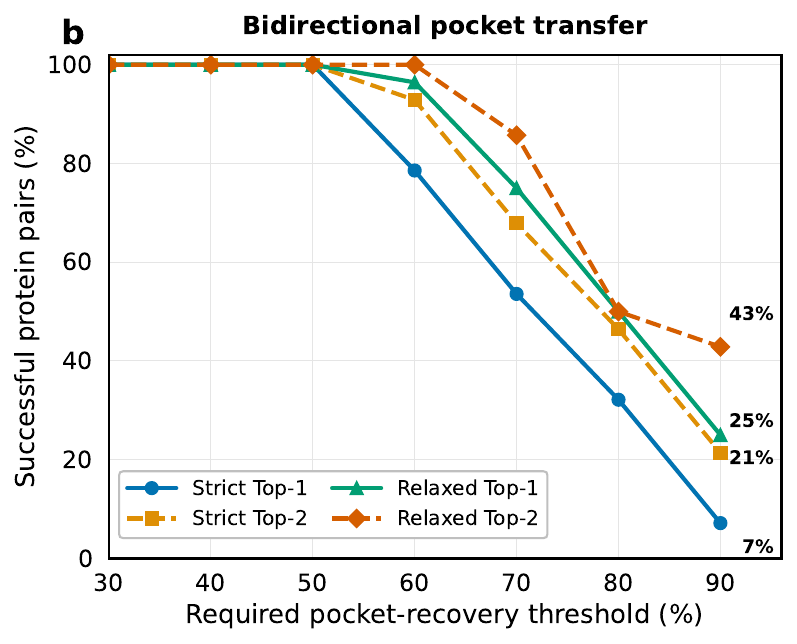}
\caption{
Threshold-dependent pocket-transfer performance for the heme-centered benchmark. The horizontal axis denotes the minimum required pocket-recovery threshold, while the vertical axis reports the percentage of successful mappings satisfying that threshold. (a) Percentage of successful directed transfers among all 56 ordered protein mappings. (b) Percentage of successful bidirectional protein pairs among the 28 unordered pairs, where a pair is counted only if the mean recovery over both mapping directions exceeds the specified threshold. Overall, relaxed-pocket evaluation consistently outperforms the strict-pocket criterion, and allowing Top-2 correspondences further improves transfer success relative to Top-1 matching.}
\label{fig:p450_pocket_transfer_thresholds}
\end{figure*}

To directly evaluate whether the computed point-to-point correspondences correctly transfer functional pocket regions between homologous proteins, we define a vertex-wise pocket-transfer score.

For each protein, we first identify the surface vertex nearest to the bound heme cofactor and use it as the center of a heme-centered pocket region. The strict pocket is defined as the set of surface vertices whose graph-geodesic distance from this vertex is at most \SI{20}{\angstrom}. This region is highly localized, covering only a small fraction of the total protein surface. For the relaxed evaluation, the target acceptance region is expanded to a graph-geodesic radius of \SI{22}{\angstrom}, allowing small localization errors near the pocket boundary while preserving the same overall functional neighborhood. 

For each directed protein pair, every source-pocket vertex is mapped to its ranked IFACE correspondences on the target surface. Under the Top-1 evaluation, only the highest-scoring correspondence is considered. Under the Top-2 evaluation, which is the primary setting used in the main analysis, a source-pocket vertex is counted as successfully transferred if either of its two highest-scoring target correspondences lies within the target pocket. Each source-pocket vertex contributes at most one successful transfer, irrespective of how many candidate correspondences fall inside the target pocket.

The pocket transfer score is defined as the fraction of source-pocket vertices that are successfully localized within the target pocket. Because the score is normalized by the number of source-pocket vertices, it measures how completely the source functional region is recovered on the target surface and permits comparison across source pockets of different sizes. The strict and relaxed target definitions are reported separately because expanding the target region changes the acceptance criterion.

To summarize transfer performance across protein pairs, we additionally define a pocket transfer hit. A directed transfer is considered successful when at least 50\% of the source-pocket vertices are correctly localized within the target pocket. This majority-recovery criterion serves as the primary operating threshold in the main analysis. 

Recovering at least half of the source pocket provides an operational indication that the corresponding heme-centered region has been substantially localized on the target surface, although the 50\% threshold should not be interpreted as a biologically validated cutoff. In addition, to assess sensitivity to the chosen stringency, we report success rates over recovery thresholds ranging from 30\% to 90\%.

Figure~\ref{fig:p450_pocket_transfer_thresholds} summarizes pocket-transfer performance across the eight P450 proteins by varying the minimum pocket transfer score required for a successful mapping. At the primary 50\% threshold, 47 of the 56 directed mappings (83.9\%) are successful under Strict Top-1, increasing to 50 mappings (89.3\%) under Strict Top-2. Under the relaxed evaluation, the corresponding success rates are 53 of 56 mappings (94.6\%) for Top-1 and 54 of 56 mappings (96.4\%) for Top-2.

For the 28 unordered protein pairs, all 28 pairs (100\%) satisfy the 50\% threshold when recovery is averaged over both mapping directions, under either the strict or relaxed pocket definition and with either Top-1 or Top-2 correspondences.
As expected, the success rates decrease monotonically as the required pocket recovery becomes more stringent. The differences between the evaluation settings become more pronounced above the 60\% threshold.

At the most stringent 90\% threshold, Strict Top-1 and Top-2 succeed for 9 and 17 of the 56 directed mappings, corresponding to 16.1\% and 30.4\%, respectively. Relaxed Top-1 and Top-2 succeed for 19 and 27 mappings, corresponding to 33.9\% and 48.2\%, respectively. Among the 28 unordered pairs, Strict Top-1 and Top-2 succeed for 2 and 6 pairs (7.1\% and 21.4\%), whereas Relaxed Top-1 and Top-2 succeed for 7 and 12 pairs (25.0\% and 42.9\%). Across the evaluated thresholds, Top-2 and relaxed-pocket evaluations yield higher success rates than Top-1 and strict-pocket evaluations, respectively, demonstrating the increased recovery obtained by considering an additional high-ranking correspondence and a slightly expanded target region.

These results should be interpreted in the context of heme-centered P450 active sites, which are typically buried or semi-buried cavities connected to solvent or membrane through access channels rather than simple exposed surface patches. In such systems, accurate transfer requires not only locating the correct heme-proximal region but also reconstructing the geometry and extent of the surrounding cavity and access channels. Recovering at least 50\% of the source-pocket vertices is therefore a useful indicator that the correct functional region has been localized, whereas higher thresholds demand increasingly faithful reconstruction of the pocket extent.

The higher performance obtained with the relaxed pocket definition is consistent with many errors occurring near pocket boundaries or small cavity offsets, but does not by itself prove that all errors are boundary-localized. Similarly, the advantage of Top-2 over Top-1 indicates that allowing an additional candidate correspondence improves recall under this evaluation, although the threshold plots alone do not establish the precise structural reason why the second-ranked candidate helps.

\section{\sb{Correspondence-quality analysis}}
\begin{figure*}[t]
    \centering

    \includegraphics[width=0.48\textwidth]{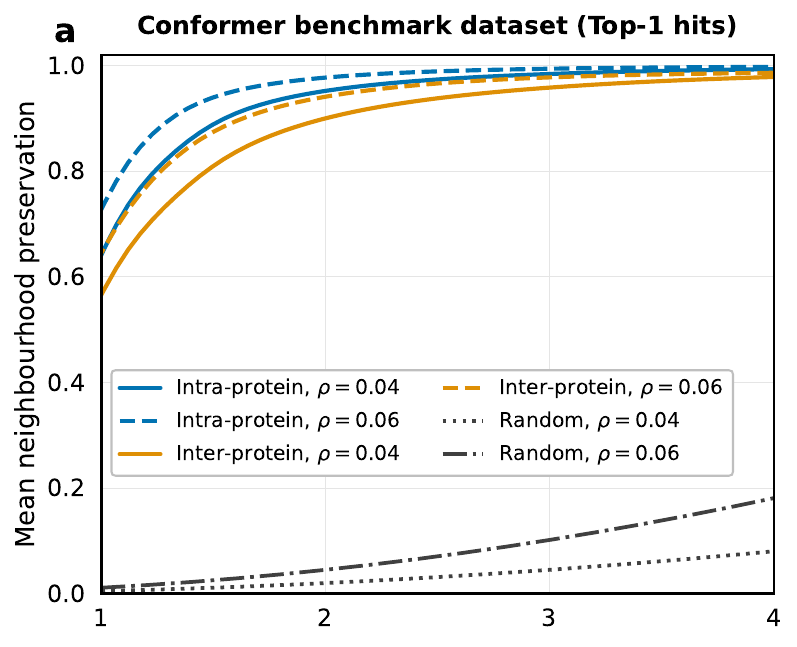}
    \hfill
    \includegraphics[width=0.48\textwidth]{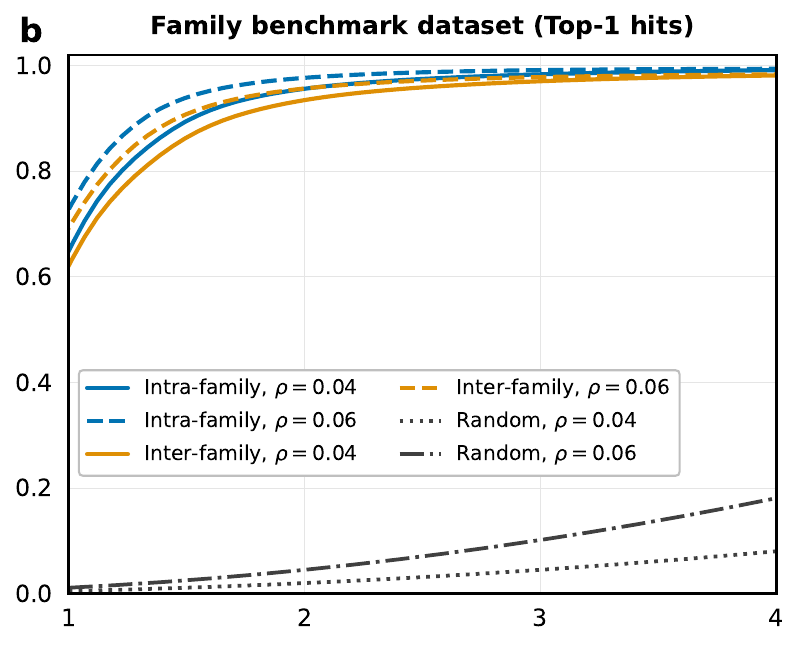}

    \vspace{0.4em}

    \includegraphics[width=0.48\textwidth]{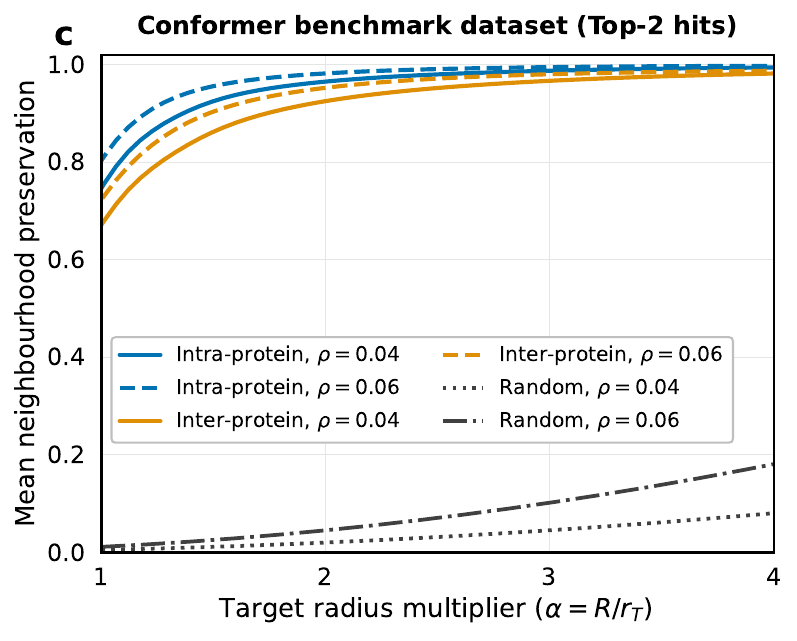}
    \hfill
    \includegraphics[width=0.48\textwidth]{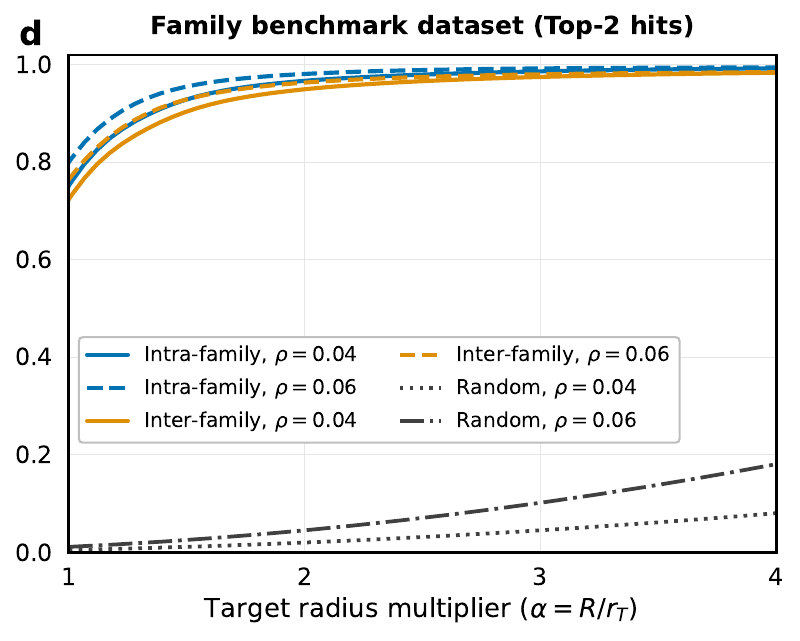}
\vspace{0.4em}

\begin{center}
\includegraphics[width=0.8\textwidth]{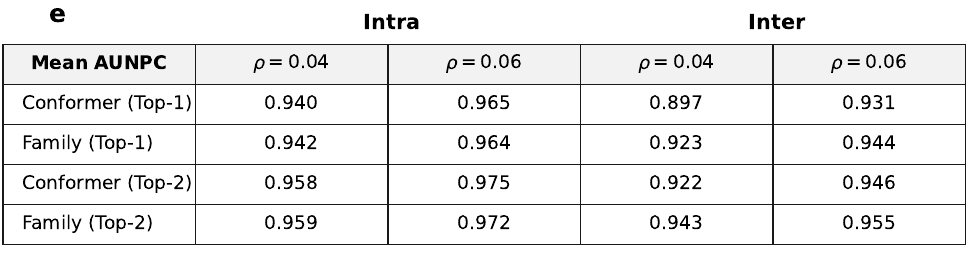}
\end{center}
\caption{
Neighborhood preservation analysis of the recovered protein surface
correspondences.
\textbf{(a)} Conformer benchmark using hard correspondences.
\textbf{(b)} Family benchmark using hard correspondences.
\textbf{(c)} Conformer benchmark using top-2 relaxed correspondences.
\textbf{(d)} Family benchmark using top-2 relaxed correspondences.
\textbf{(e)} Mean Area Under the neighborhood Preservation Curve (AUNPC)
for the four experimental settings.
Panels (a)--(d) show the mean neighborhood preservation as a function of the
target radius multiplier $\alpha$ for $\rho=0.04$ and $\rho=0.06$.
}
    \label{fig:ngnps_comparison}
\end{figure*}

To assess the quality of the recovered correspondences, we evaluate neighborhood preservation using geodesic neighborhoods (Fig.~\ref{fig:ngnps_comparison}). For each source vertex, a neighborhood is defined by a geodesic ball of radius $r_S=\rho\sqrt{A_S}$, where $A_S$ is the total surface area of the source mesh. The corresponding target neighborhood is centered at the target vertex to which the source vertex is mapped and has radius $R=\alpha r_T$, where $r_T=\rho\sqrt{A_T}$ and $A_T$ is the total surface area of the target mesh.

The neighborhood preservation score is defined as the fraction of mapped source neighbors that lie within the target geodesic ball, averaged over all source vertices. We report this quantity as a function of the target radius multiplier $\alpha$ for $\rho=0.04$ and $\rho=0.06$, corresponding to neighborhoods containing approximately 12 and 30 vertices on average.

At $\alpha=1$, the source and target neighborhoods are defined at the same area-normalized geodesic scale. Since the source and target meshes have comparable sampling densities and similar numbers of vertices, they contain approximately the same number of vertices on average. Thus, $\alpha=1$ provides the natural reference point for neighborhood preservation, while $\alpha>1$ measures the additional target radius required to recover the mapped neighborhood.

To summarize the entire curve with a single scalar, we compute the Area Under the neighborhood Preservation Curve (AUNPC) by integrating the neighborhood preservation curve over $\alpha\in[1,4]$ and normalizing by the integration interval. We report results using both hard correspondences, obtained by assigning each source vertex to its highest-probability target vertex, and a top-2 relaxed evaluation, in which a neighboring source vertex is considered preserved if either of its two highest-probability target candidates lies within the target geodesic ball.

Figure~\ref{fig:ngnps_comparison} demonstrates that IFACE recovers correspondences that preserve local surface neighborhoods across both the conformer and family benchmarks. In all cases, neighborhood preservation increases monotonically with the target radius multiplier $\alpha$ and approaches unity as the target neighborhood expands, indicating that mapped neighbors remain spatially coherent under the recovered correspondences.

As expected, intra-protein and intra-family pairs consistently exhibit higher neighborhood preservation than inter-protein and inter-family pairs, respectively, particularly around the natural operating point $\alpha=1$, where the source and target neighborhoods are defined at the same normalized geodesic scale. This separation indicates that IFACE produces more locally consistent correspondences for same-protein conformers and proteins from the same family than for unrelated proteins, while the gradual convergence of the curves at larger $\alpha$ reflects the expected recovery of neighborhoods as the target radius is expanded.

The top-2 relaxed evaluation consistently improves neighborhood preservation over the hard assignment on both benchmarks, indicating that the soft correspondence matrix retains additional geometrically consistent candidate matches beyond the single best assignment. Nevertheless, the improvement is modest, suggesting that the highest-probability correspondence already captures most of the local neighborhood structure recovered by IFACE.

These observations are summarized quantitatively in Fig.~\ref{fig:ngnps_comparison}(e), which reports the mean Area Under the neighborhood Preservation Curve (AUNPC). Across all evaluated settings, AUNPC values exceed 0.89 and consistently increase under the top-2 relaxed evaluation.

For example, on the family benchmark, the mean AUNPC increases from 0.942 to 0.959 for $\rho=0.04$ and from 0.964 to 0.972 for $\rho=0.06$ for intra-family pairs, while the corresponding inter-family values increase from 0.923 to 0.943 and from 0.944 to 0.955, respectively.

Similarly, on the conformer benchmark, the mean AUNPC increases from 0.940 to 0.958 ($\rho=0.04$) and from 0.965 to 0.975 ($\rho=0.06$) for intra-protein pairs, while the corresponding inter-protein values increase from 0.897 to 0.922 and from 0.931 to 0.946, respectively.

Overall, these consistently high AUNPC values indicate that IFACE recovers correspondences with strong local geometric consistency, while the gains achieved by the top-2 evaluation indicate that the soft correspondence matrix retains additional plausible local matches beyond the highest-probability assignment.

\vspace{1cm}

\section{\sb{Scaling}}
\begin{figure*}[t]
    \centering

    \includegraphics[width=0.48\textwidth]{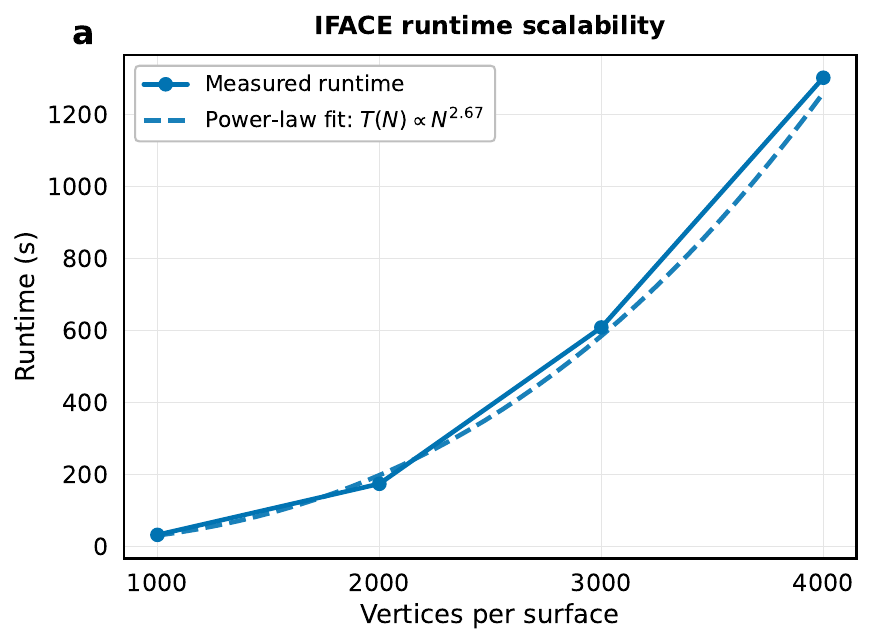}
    \hfill
    \includegraphics[width=0.48\textwidth]{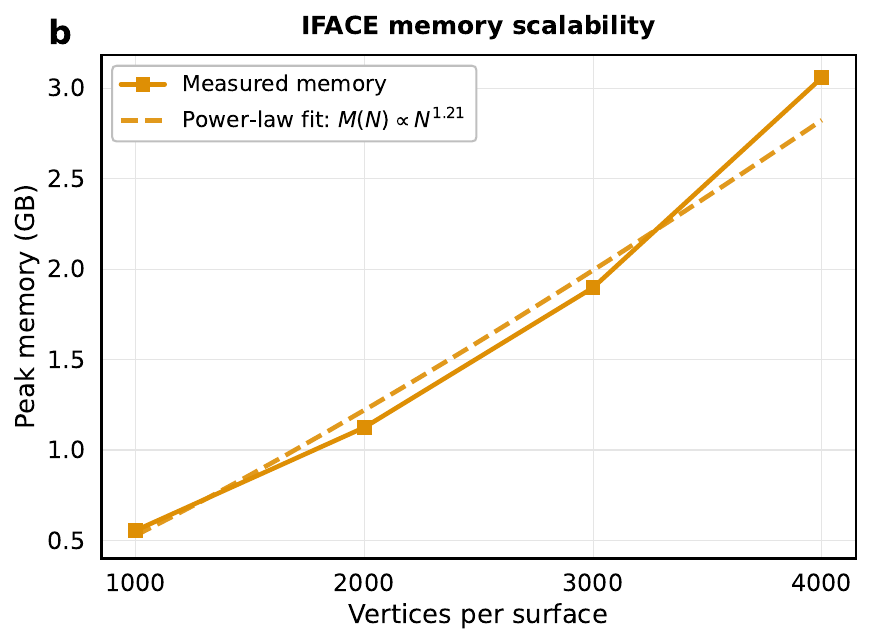}

    \caption{
    Scalability analysis of IFACE with increasing mesh resolution.
    A representative protein pair was uniformly downsampled to different mesh
    resolutions while keeping all algorithmic parameters fixed.
 \textbf{Left:} Total runtime of the complete IFACE pipeline, including coupling optimization and distance calculation.
\textbf{Right:} Peak resident memory usage during execution.
Solid curves show the measured values, while dashed curves represent least-squares power-law fits of the form $y = CN^{p}$, illustrating the empirical scaling of IFACE with mesh resolution (number of vertices per surface).
    }
    \label{fig:iface_scalability}
\end{figure*}

To evaluate the computational scaling of IFACE, we measured the runtime and peak memory consumption of the complete pipeline for a representative protein pair while varying the mesh resolution. Both the source and target surfaces were uniformly downsampled to approximately 1,000, 2,000, 3,000, and 4,000 vertices, while all algorithmic parameters and optimization settings were held fixed. Figures~\ref{fig:iface_scalability}(a) and \ref{fig:iface_scalability}(b) show the total runtime and peak resident memory usage, respectively, as functions of the number of vertices per surface. The dashed curves denote least-squares power-law fits of the form ($y=CN^{p}$). Over the range of mesh resolutions considered, the fitted relations were ($T(N)\approx3.17\times10^{-7}N^{2.67}$) for runtime and ($M(N)\approx1.24\times10^{-4}N^{1.21}$) for peak memory usage, where runtime is measured in seconds and memory in gigabytes.

\section{\sb{Discussion}}

Protein surfaces are naturally represented as coupled geometric--chemical objects, and their similarity is most directly defined through correspondence between surfaces. IFACE implements this principle by constructing an optimal-transport coupling, yielding both an explicit soft correspondence and a symmetric geometric--chemical distance. Similarity is therefore determined by how geometry and physicochemical fields align across complete surfaces, rather than by comparing them separately.

Across the proteins examined here, surface-based distances distinguished intra-protein conformational variability from inter-protein divergence more effectively than global fold-based similarity measures. Incorporating the coupled organization of surface geometry and physicochemical fields reduced the overlap between the intra-protein and inter-protein distance distributions. The stronger performance of JSD in this binary identity test shows that aggregate surface-feature distributions can be sufficient when the task is simply to distinguish one protein from another. This indicates that protein identity is reflected not only in the global fold but also in the spatial arrangement of geometric and physicochemical properties across the molecular surface.

At the family level, IFACE coherently organized proteins from six families and separated the outlier. Geometric distances captured shared surface shape and curvature organization, whereas chemical distances captured complementary physicochemical similarity. Their combination produced stable separation between intra-family and inter-family protein pairs. The contribution of mean curvature further indicates that extrinsic surface organization contains information not fully represented by intrinsic geometry or chemical fields alone.

Compared with the baseline methods, IFACE provided stronger family-level organization and separation of intra-family and inter-family pairs. The weaker performance of JSD-based distances indicates that similarity between marginal feature distributions is insufficient when the spatial arrangement of surface properties is discarded. Likewise, the Laplace--Beltrami spectral distance captures global intrinsic geometry but does not incorporate physicochemical organization. The learning-based MaSIF and SurfaceID descriptors also showed weaker family-level discrimination on the present dataset. This may reflect their design for task-specific local surface comparisons rather than for defining a general distance between complete protein surfaces. These comparisons support the importance of jointly evaluating geometry and physicochemical fields through explicit correspondence.

The results indicate that family-level relationships are reflected in coupled surface organization. Proteins belonging to the same family retained related geometric and physicochemical surface patterns despite sequence variation and differences in species of origin. The present analysis does not establish a direct mechanistic relationship between the IFACE distance and biological function. Nevertheless, the consistent family-level separation suggests that coupled surface organization may provide a useful indicator of functional relatedness.

An important advantage of IFACE is that the correspondence itself is an output of the comparison. The resulting mapping supports both global analyses, including classification and clustering, and local analyses of corresponding pockets and surface regions. The neighborhood-preservation and pocket-transfer analyses further demonstrate that the correspondence preserves coherent local surface organization. IFACE therefore provides both a global similarity measure and an interpretable mapping without requiring task-specific supervision.

Several limitations warrant consideration. The present implementation uses a limited set of scalar surface fields, and the benchmarks contain a limited number of protein families and conformational ensembles. Broader validation across more diverse proteins and functional surface regions will be required to establish generality. 
Performance remained comparable when the surface resolution was reduced from 3,000 to 1,000 vertices, indicating that the principal geometric--chemical organization is retained at substantially lower computational cost.

Nevertheless, the computational cost increases with surface resolution, motivating future development of sparse, multiscale, or accelerated transport formulations. Extensions to vector- or tensor-valued fields and alternative surface representations may further refine the correspondence and distance definitions. Future studies can test whether the same framework supports large-scale retrieval and comparison of binding pockets, interaction interfaces, and other functionally annotated surface regions.

Overall, IFACE formulates protein-surface similarity as a coupled geometric--chemical comparison mediated by explicit correspondence between surfaces. By providing both a global distance and an interpretable local mapping, it offers a physically explicit, mathematically principled, and unsupervised framework for comparing functional protein surfaces.

\vspace{1cm}

\section{\sb{Methods}}

In this section, we describe how distances between protein surfaces are constructed within the IFACE framework. The key step is the identification of a soft correspondence between two surfaces, encoded in an optimal coupling matrix that relates their surface vertices.

\nsb{Optimal Coupling Matrix.}
We consider two protein surfaces, $S_\alpha$ and $S_\beta$, represented as triangulated meshes and endowed with $m$ surface feature fields, denoted $f^\alpha_m$ and $f^\beta_m$. Our central objective is to relate these two surfaces in a manner that respects \emph{both} their geometric structure and surface-field organization. To achieve this, we introduce a probabilistic, or \emph{soft}, correspondence between the surfaces, encoded by a \emph{coupling matrix} $\mathbf{P}$. Each matrix element $P_{ij}$ quantifies the degree to which vertex $i$ on $S_\alpha$ corresponds to vertex $j$ on $S_\beta$. By avoiding a rigid one-to-one assignment, this formulation permits a flexible comparison of surface regions and naturally accommodates structural variability and chemical heterogeneity.

The coupling matrix is constrained by prescribed marginal distributions over the vertices of each surface, $\rho^{\alpha}$ and $\rho^{\beta}$, such that $\sum_j P_{ij} = \rho_i^{\alpha}$ and $\sum_i P_{ij} = \rho_j^{\beta}$. These marginals assign weights to surface elements based on local surface area and the values of the surface feature fields. They therefore act as nonuniform vertex weights in the correspondence constraints. The explicit construction of these distributions is described in the Supplementary Information.

The optimal coupling matrix $\mathbf{P}$ is determined by balancing two complementary contributions. The first is a \emph{field term}, which compares the surface feature fields defined on the two surfaces. The second is a \emph{structural term}, which enforces geometric compatibility between surface shapes. The optimal correspondence emerges from reconciling these two contributions under the marginal constraints above, yielding a coupling that reflects both field similarity and geometric coherence. We now define these two terms explicitly.

\nsb{Field Term.}
The field term $\bm{\mathcal{F}}$ quantifies the mismatch between feature fields on the two surfaces. For optimization purposes, it is defined as
\begin{equation*}
\begin{aligned}
\bm{\mathcal{F}}(S_\alpha, S_\beta)
&= \sum_{i,j,m} P_{ij}
\left[
\frac{ f^{\alpha}_{m,i} - f^{\beta}_{m,j} }{ \sigma^{m} }
\right]^2 ,
\end{aligned}
\end{equation*}
where $\sigma^{m}$ denotes the standard deviation of the $m^{\text{th}}$ feature field, computed using values from both surfaces. This normalization places all fields on a comparable scale and promotes the matching of vertices with similar surface-field values.

\nsb{Structural Term.}
The structural term enforces consistency between the intrinsic geometries of the two surfaces. To this end, we construct for each surface $S$ its geodesic distance matrix $D^S_{ij}$ on the unit-normalized mesh, supplemented by a smoothing kernel $H^S_{ij}$ that incorporates global geometric information. These are combined as
$G^S_{ij} = D^S_{ij} + \epsilon_g\, H^S_{ij}$, with $S \in \{S_\alpha, S_\beta\}$.
Here, the smoothing kernel is 
$H^S_{ij} = \sum_k K^S_{ik}\,K^S_{kj}$, with
$K^S_{lp} = \exp\![-(D^S_{lp})^2/(2\sigma_S^2)]$.
Thus, $H^S$ encodes smoothed, long-range geometric correlations derived from the geodesic matrix. 

Using these definitions, the structural discrepancy between surfaces $S_\alpha$ and $S_\beta$ is given by
\begin{equation*}\Label{Eq:structural}
\bm{\mathcal{S}}(S_\alpha, S_\beta)
= \sum_{i,j,k,l}
P_{ik}\, P_{jl}
\left( G^{\alpha}_{ij} - G^{\beta}_{kl} \right)^2~.
\end{equation*}
In our analysis, $\sigma_S$ is chosen as a fixed fraction of the maximum geodesic distance, $\sigma_S = 0.1\,D^S_{\max}$, and the weight $\epsilon_g$ is set to $0.1$ to ensure that the global contribution remains a perturbative correction of the geodesic structure. In the limit $\epsilon_g = 0$, this term reduces to the classical Gromov--Wasserstein objective, with $G^S_{ij} \to D^S_{ij}$~\cite{memoli2011gromov, vayer2018fused}.

\nsb{Optimization.}
To determine the coupling matrix, we minimize the following entropic-regularized objective function:
\begin{align}
P_{ij}^{*}
=  \arg \min_{P_{ij}}
\Bigl[
(1 - \lambda)\,&\bm{\mathcal{F}}(S_\alpha, S_\beta)  \nonumber \\
+  \lambda\,&\bm{\mathcal{S}}(S_\alpha, S_\beta)
+ \epsilon \sum_{i,j} P_{ij}\log P_{ij}
\Bigr],
\end{align}
where the minimization is subject to the constraints
$\sum_j P_{ij} = \rho_{i}^{\alpha}$ and
$\sum_i P_{ij} = \rho_{j}^{\beta}.$
The entropic term
$-\epsilon \sum_{i,j} P_{ij}\log P_{ij}$
plays a distinct and complementary role to the geometric regularization controlled by $\epsilon_g$. While $\epsilon_g$ smooths the \emph{metric structure} entering the structural term, the entropy promotes diffuseness of the coupling itself, suppressing spurious sharp correspondences and stabilizing the optimization landscape. In practice, this regularization improves convergence and prevents the solution from becoming trapped in shallow local minima.

To find the optimal coupling, we proceed in several stages. We first obtain an initial alignment of the meshes using a combination of rigid and feature-field-aware non-rigid alignment, based on mean curvature, hydrophobicity, hydrogen bonding, and electrostatic potential. This alignment is converted into an initial coupling matrix using a Gaussian kernel followed by Sinkhorn normalization~\cite{sinkhorn1967diagonal,cuturi2013sinkhorn} to enforce the marginal constraints. The resulting coupling serves as a warm start for minimizing the entropically regularized objective. 
Finally, the entropically regularized solution is used to initialize a second optimization with $\epsilon=0$, yielding the final coupling matrix $P_{ij}^{*}$.
Full algorithmic details are provided in the Supplementary Information.

We next describe how distances between protein surfaces are constructed from the resulting coupling matrix.

\nsb{IFACE Distance.}
We construct distances between protein surfaces by leveraging the optimal coupling to compare both surface feature fields and geometric structure. The guiding principle is to use only the most reliable portions of the correspondence: for each vertex, we retain the two most confidently mapped vertices on the opposing surface. Feature values are transported across the coupling, evaluated at the mapped vertices, and compared using the $\ell^1$ norm, \ie, absolute value. The choice of the $\ell^1$ norm is deliberate: it suppresses the influence of outliers that can arise from discretization effects, where feature values may vary unevenly across meshes of differing resolution, thereby yielding a more robust estimate of feature-field dissimilarity.

To ensure symmetry, all comparisons are performed \emph{bidirectionally}. In addition, prior to computing distances, feature fields are mildly regularized by local smoothing. For each vertex $i$, the field value is updated using its neighborhood according to
\[
\tilde f_i \equiv f_i + \epsilon_f \sum_k \exp\!(-D_{ik}/\beta_i)\,f_k,
\]
where $D$ is the geodesic distance matrix and $\beta_i$ is chosen as the characteristic distance of the 2-ring neighborhood of vertex $i$. Throughout this work, we fix $\epsilon_f = 0.1$. 
In the following expressions, $\tilde f_i$ denotes the smoothed field.

Using these smoothed fields, the feature-field distance $D_{\text{field}}$ is defined as
\begin{equation*}
d_{\alpha\rightarrow\beta}
=\sum_{i, q} \,\omega_i^q\,\big\lVert\tilde f^{\alpha}_{i}-\tilde f^{\beta}_{y^q_{i}}\big\rVert_{l^1}, \quad
d_{\beta\rightarrow\alpha}
=\sum_{j, r} \,\eta_j^r\,\big\lVert \tilde f^{\beta}_j-\tilde f^{\alpha}_{x_j^{r}}\big\rVert_{l^1},
\end{equation*}\vspace{-0.8\baselineskip}
\begin{equation*}
D_{\text{field}}(S_\alpha,S_\beta)
=\frac{d_{\alpha\rightarrow\beta}+d_{\beta\rightarrow\alpha}}{2},
\Label{eq:Dfield}
\end{equation*}
where $\omega_i^q$ and $\eta_j^r$ are confidence weights associated with the vertex mappings
$i \rightarrow y_i^q \in \{y^j\}$ and $j \rightarrow x_j^r \in \{x^i\}$, respectively. These weights are normalized such that
$\sum_{q=1}^2 \omega_i^q = 1$ and $\sum_{r=1}^2 \eta_j^r = 1$, ensuring that only the relative confidence between the two dominant correspondences enters the distance.

The structural distance is constructed analogously using a
bidirectional comparison of the structural matrices
$G^\alpha$ and $G^\beta$, which are computed from the original
protein meshes rather than the unit-normalized meshes. It is defined as
\begin{align}
D_{\text{structure}}(S_\alpha,S_\beta)
=\frac{1}{\sqrt{2}}\Big(
\sum_{i,j,q,p}
\omega_i^q \omega_j^p
\big\lVert
G^{\alpha}_{ij}
-
G^{\beta}_{y^q_i y^p_j}
\big\rVert^2
+ \nonumber \\
\sum_{k,l,r,s}
\eta_k^r \eta_l^s
\big\lVert
G^{\alpha}_{x_k^r x_l^s}
-
G^{\beta}_{kl}
\big\rVert^2
\Big)^{1/2}.
\end{align}

All notation follows the definitions introduced above. This construction yields a symmetric structural distance that reflects discrepancies in intrinsic geometry while remaining tightly coupled to the most confident regions of the surface correspondence.

To combine the geometric and chemical information into a unified surface-comparison distance, we first normalize each distance component before aggregation. The normalization bounds are computed from a reference dataset of 130 proteins spanning both the protein-family and conformer benchmarks. This reference dataset provides a broad distribution of distances, allowing the different components to be placed on comparable numerical scales and preventing any component from dominating solely because of its magnitude.

For each surface scalar field \(m\), the range-normalized field distance is defined as

\begin{equation*}
\bar{D}_m(S_\alpha,S_\beta)
=
\frac{
D_m(S_\alpha,S_\beta)
}{
D_{\max,m}-D_{\min,m}
},
\Label{eq:field_normalization}
\end{equation*}

where \(D_m(S_\alpha,S_\beta)\) is the bidirectional field distance defined above, and \(D_{\min,m}\) and \(D_{\max,m}\) are the minimum and maximum values, respectively, of that distance component in the reference dataset. 
The empirical range $D_{\max,m}-D_{\min,m}$ is used as a scale factor; the reference minimum is not subtracted, thereby preserving zero raw distance.
The structural distance is normalized independently as

\begin{equation*}
\bar{D}_{\mathrm{struct}}(S_\alpha,S_\beta)
=
\frac{
D_{\mathrm{struct}}(S_\alpha,S_\beta)
}{
D_{\max,\mathrm{struct}}
-
D_{\min,\mathrm{struct}}
},
\Label{eq:structural_normalization}
\end{equation*}

where \(D_{\min,\mathrm{struct}}\) and \(D_{\max,\mathrm{struct}}\) are the corresponding reference bounds for the structural distance.

Because mean curvature is a geometric scalar field, the geometric distance is defined as the sum of the normalized structural distance and the normalized mean-curvature field distance,

\begin{equation*}
D_{\mathrm{geom}}(S_\alpha,S_\beta)
=
\bar{D}_{\mathrm{struct}}(S_\alpha,S_\beta)
+
\bar{D}_{\mathrm{MC}}(S_\alpha,S_\beta),
\Label{eq:geometric_distance}
\end{equation*}

where \(\bar{D}_{\mathrm{MC}}\) denotes the range-normalized bidirectional distance between the mean-curvature fields.

The chemical distance is defined as the sum of the normalized physicochemical field distances,

\begin{equation*}
D_{\mathrm{chem}}(S_\alpha,S_\beta)
=
\sum_{m=1}^{M_{\mathrm{chem}}}
\bar{D}^{\,\mathrm{chem}}_m(S_\alpha,S_\beta),
\Label{eq:chemical_distance}
\end{equation*}

where \(M_{\mathrm{chem}}\) is the number of physicochemical surface fields. In the present formulation, these fields are electrostatic potential, hydrogen-bond propensity, and hydrophobicity.

Finally, the IFACE distance is defined as the sum of the geometric and chemical distances,

\begin{equation}
D_{\mathrm{IFACE}}(S_\alpha,S_\beta)
=
D_{\mathrm{geom}}(S_\alpha,S_\beta)
+
D_{\mathrm{chem}}(S_\alpha,S_\beta).
\label{eq:iface_distance}
\end{equation}

\bigskip
\nsb{Data and Software Availability}

\noindent The dataset used for the analysis can be freely downloaded from Zenodo:
\href{https://doi.org/10.5281/zenodo.21760918}{https://doi.org/10.5281/zenodo.21760918}.
A working implementation of the IFACE framework, including surface preprocessing, coupling computation, and distance evaluation, is available on GitHub:
\href{https://github.com/ihswami/iface}{https://github.com/ihswami/iface.}

\section*{Author Contributions}
H.S. conceived the differential-geometric approach underlying IFACE and developed the IFACE methodology presented in this work, including its mathematical formulation and computational implementation, wrote the software, performed the analyses and validation, curated the data, prepared the original draft, and contributed extensively to its revision. J.M.M., J.-P.E., and T.T. conceived the broader project of mapping and comparing protein surfaces. J.M.M. developed software for dataset processing and method evaluation and contributed to data curation, supervision, and writing. J.-P.E. contributed to mathematical and theoretical development, interpretation of the results, scientific supervision, and writing. T.T. guided and supervised the scientific development of the project and contributed to interpretation and writing.\\                                  

\section*{Acknowledgments}
This work was supported by the National Research Foundation of Korea under Grant No. NRF-RS-2025-00573354, the InnoCORE Bio-MAX program, and the U.S. Office of Naval Research under Award No. N00014-26-1-2401. J.-P.E. was partially supported by SwissMAP.

\newpage

\bibliography{main}

\clearpage

\noindent\sb{Supplementary Materials}

\section{\sb{The General IFACE Framework}}
Let $S_\alpha$ and $S_\beta$ be two  protein surfaces, each equipped with a collection of fields $F^{\alpha} = \{f^\alpha_m\}$ and $F^{\beta} = \{f^\beta_m\}$, where each $f^\alpha_m: S_\alpha \rightarrow \mathbb{R}^{n_m}$ and $f^\beta_m: S_\beta \rightarrow \mathbb{R}^{n_m}$ may represent a scalar field ($n_m = 1$), a vector field, or a tensor field. Although protein surfaces can be described more richly using vector and tensor fields (\eg, electric fields, stress–strain~\cite{mitchell2016}, and polarizability tensors), they are still studied predominantly with scalar fields. Therefore, in this work, we restrict our analysis to \textbf{scalar} fields on protein surfaces—such as electrostatic potential, hydrogen bonding propensity, and hydrophobicity. Our goal is to define a distance metric $D(S_\alpha, S_\beta)$ that jointly captures both geometric and field-level differences.\\

\nsb{From Diffeomorphic Mapping to Probabilistic Soft Mapping for Field Transport.}
Two surfaces are considered equivalent when there exists a diffeomorphism that maps one to the other while leaving all transported  fields unchanged, \ie,

\begin{equation*}
\begin{aligned}
S_\alpha \sim S_\beta
\;\Longleftrightarrow\;
\exists\,\varphi\in\mathrm{Diff}(S_\alpha,S_\beta)
\ \text{such that}
\\
\varphi^{*}g_\beta=g_\alpha
\quad\text{and}\quad
\varphi^{*}F^\beta=F^\alpha .
\end{aligned}
\end{equation*}
Here, $\varphi^{*}F^{\beta}$ denotes the family of fields on $S_\alpha$ obtained by pulling back the fields in $F^{\beta}$, and $\varphi^{*}g_\beta$ denotes the metric on $S_\alpha$ obtained by pulling back $g_\beta$, via the diffeomorphism $\varphi\colon S_\alpha\to S_\beta$.

Equivalently, the similarity between the pullback fields and the native fields
can be quantified using $L^p$ norms. We define the pullback distance between
families of scalar fields as the $\ell^1$-sum of the componentwise $L^p$
distances:
\begin{equation*}\Label{eq:field_lp}
d_{F}^{(p)}(F^\alpha,F^\beta;\varphi)
:=
\sum_m
\left(
\int
\left|
f_\alpha^m - \varphi^{*}f_\beta^m
\right|^{p}
\, d\mu^\alpha
\right)^{\frac{1}{p}}.
\end{equation*}
Here, $p \ge 1$ denotes the exponent of the $L^p$ norm.
 $d\mu^\alpha$ denotes the surface-area measure on the  surface ($S_{\alpha}$) and it is given by $d\mu^\alpha \equiv \sqrt{|g_{\alpha}(x)|}d^2x$, where $|g_{\alpha}(x)|$ is the determinant of the metric tensor on the surface $S_\alpha$.  Further, to compare the internal geometric structure, we compare the
metric tensor fields using the pushforward map $\varphi_{*}$:
\begin{equation*}\Label{eq:lq_metric}
d_G^{(q)}(g_\alpha,g_\beta;\varphi)
:=
\left(
\int_{S_\beta}
\left\|
\varphi_{*}g_{\alpha}-g_{\beta}
\right\|^{q}
\, d\mu^\beta
\right)^{\frac{1}{q}}.
\end{equation*}
Here $d\mu^\beta$ denotes the surface area measure on $S_\beta$,
$q \ge 1$ is the exponent for the geometric discrepancy, and
$\|\cdot\|$ denotes a pointwise tensor norm. The exponents $p$ and $q$ control the sensitivity of the field
and geometric discrepancies, respectively.
Furthermore, similar expressions can be obtained for the inverse mapping $\varphi^{-1}$.
We can now define a distance measure, $D(S_\alpha,S_\beta)$, using the diffeomorphism mapping, as follows

\begin{equation}\label{eq:hard_distance}
\begin{aligned}
D(S_\alpha, \,S_\beta)
=\;
&\sum_m \zeta_m
\Bigg(\frac{1}{2}\Bigl(
\int_{S_\alpha}
| f_\alpha^m-\varphi^{*}f_\beta^m |^{p}\,d\mu^\alpha
\\
&\qquad\qquad
+\int_{S_\beta}
| f_\beta^m-\varphi^{-1*}f_\alpha^m |^{p}\,d\mu^\beta
\Bigr)\Bigg)^{\frac{1}{p}}
\\[4pt]
&+\;
\eta
\Bigg(\frac{1}{2}\Bigl(
\int_{S_\beta}
\|\varphi_* g_{\alpha}-g_{\beta}\|^{q}\,d\mu^\beta
\\
&\qquad\qquad
+\int_{S_\alpha}
\|\varphi^{-1}_* g_{\beta}-g_{\alpha}\|^{q}\,d\mu^\alpha
\Bigr)\Bigg)^{\frac{1}{q}}.
\end{aligned}
\end{equation}
Here $\zeta_m$ and $\eta$ denote weighting coefficients for the
$m^{\text{th}}$ scalar field and the metric tensor discrepancy,
respectively. These weights control the relative importance of the
terms and provide normalization so that the resulting quantities are
dimensionless or measured in the same units.
We note that the above metric (Eq. \ref{eq:hard_distance}) is clearly symmetric under the exchange, $S_\alpha
\leftrightarrow S_\beta$,  and furthermore, it can be shown that it
satisfies the properties of a metric.\\

\nsb{Probabilistic Soft Mapping.}
We have described a distance metric that assumes the existence of a
diffeomorphism between the surfaces. However, protein surfaces are often corrupted by noise and limited resolution and an \emph{exact} diffeomorphism rarely exists, even between two conformers of the \emph{same} protein.   
Therefore, rather than searching for a single hard map $\varphi : S_{\alpha} \to S_{\beta}$, we relax the correspondence to a probability distribution on the product space $S_{\alpha} \times S_{\beta}$.  
In other words, we allow each point $x \in S_{\alpha}$ to match several points $y \in S_{\beta}$ with varying likelihoods. In addition, this approach has the advantage that, instead of producing a continuous map, it can probabilistically map different parts of proteins that have similar patches. For example, two proteins may share only a pocket, while the other parts of the proteins differ from each other.

Formally, we introduce a coupling $ p(x \, , \,  y)$.
\begin{equation*}
\begin{aligned}
p(x \, , \,  y)  &  \in \mathcal{P}(S_\alpha \times S_\beta), \,\,\, p(x , y) \geq 0, \, \\\int_{S_\alpha} p(x &\, , \,  y) d\mu^\alpha(x) =\rho^\beta(y), \, \quad \int_{S_\beta} p(x \, , \,  y)d\mu^\beta(y) =\rho^\alpha(x), 
\end{aligned}   
\end{equation*}
where $\rho^\alpha(x)$ and $\rho^\beta(y)$ are the marginal distributions on the respective spaces.  These marginal distributions represent, in a physical sense, the probability--or weight--with which each point on the surface should be considered when being mapped to the other. They can be chosen according to prior knowledge about which regions of one surface are more likely to be present and have a correspondence to regions on the other. They satisfy the normalization conditions:
\begin{equation*}
    \int_{S_\alpha} \rho^\alpha(x) \, d\mu^\alpha(x) = 1, \quad
    \int_{S_\beta} \rho^\beta(y) \, d\mu^\beta(y) = 1,
\end{equation*}
which ensure that $\rho^\alpha$ and  $\rho^\beta $ are valid probability density functions on $S_\alpha$ and $S_\beta $, respectively. 
Incorporating these into the Eq. \ref{eq:hard_distance} leads to a distance which we call the soft distance: \\
$D_{\mathrm{soft}}(S_\alpha,S_\beta)
=$
\begin{equation*}\Label{eq:soft_distance}
\begin{aligned}
&\sum_m \zeta_m
\Bigg(
\frac{1}{2}\int_{S_\alpha}\!\!\int_{S_\beta}
\bigl| f_\alpha^m(x) - f_\beta^m(y) \bigr|^{p}
\, p(x,y)\, d\mu^\alpha(x)\, d\mu^\beta(y) \\
&+\; (S_\alpha \leftrightarrow S_\beta)
\Bigg)^{\frac{1}{p}}+\;
\eta
\Bigg(
\frac{1}{2}\int_{S_\alpha}\!\!\int_{S_\beta}
\bigl\| g_{\alpha}(x) - g_{\beta}(y) \bigr\|^{q} 
\, p(x,y)\, \\
& d\mu^\alpha(x)\, d\mu^\beta(y) \; + (S_\alpha \leftrightarrow S_\beta)
\Bigg)^{\frac{1}{q}}
.
\end{aligned}
\end{equation*}
We note that comparing metric tensors requires a transformation that
involves the pushforward (\ie, transporting) of tangent spaces
between the surfaces. This, in turn, necessitates the introduction of
additional structure into our framework and thereby increases
complexity. To minimize this complexity and make the comparison more
tractable, we relax the notion of local distance---originally encoded
by the metric tensor---to a more global measure on the surface by
using pairwise geodesic distances. Incorporating this into our
formulation, the distance becomes 
\begin{equation}\label{eq: soft_global_distance}
\begin{aligned}
D^{\text{soft}}(S_\alpha, S_\beta) = D_\text{field}^{\text{soft}}(S_\alpha, S_\beta) +  D_\text{geom}^{\text{soft}}(S_\alpha, S_\beta)~,
\end{aligned}
\end{equation}
where 
\begin{equation*}
\begin{aligned}
D_\text{field}^{\text{soft}}(S_\alpha, S_\beta) :=\sum_m \zeta_m
\Bigr(
\frac{1}{2}\int_{S_\alpha}\!\!\int_{S_\beta}
\bigl| f_\alpha^m(x) - f_\beta^m(y) \bigr|^{p} \\
\, p(x,y)\, d\mu^\alpha(x)\, d\mu^\beta(y) + (S_\alpha \leftrightarrow S_\beta) 
\Bigl)^{\frac{1}{p}} ~,
\end{aligned}
\end{equation*}
\begin{equation*}
\begin{aligned}
D_{\text{geom}}^{\text{soft}}(S_\alpha, S_\beta)
=\;
\eta\Bigg(
\int_{S_\alpha^2}\!\!\int_{S_\beta^2}\frac{1}{2}
\bigl| d_{\alpha}(x, x') - d_{\beta}(y, y') \bigr|^{q}
\, \\p(x,y) p(x',y') d\mu^\alpha(x)\, d\mu^\alpha(x')\,
d\mu^\beta(y)\, d\mu^\beta(y') \\+ (S_\alpha \leftrightarrow S_\beta) 
\Bigg)^{\frac{1}{q}}~,
\end{aligned}
\end{equation*}
where $d_{\alpha}(x, x')$ and $ d_{\beta}(y, y')$ denote the pairwise
geodesic distances on $S_\alpha$ and $S_\beta$, respectively. Note
that, since we are comparing pairs of points, we include an additional
soft correspondence term $p(x', y')$ in the
integral. Further, Eq.~\ref{eq: soft_global_distance} is
symmetric under exchange $S_\alpha \leftrightarrow S_\beta$. As we are going
to use meshes, we add a term which further strengthens the global structure 
\begin{equation*}
h_{S}(x,x') = \int_{\Omega} k_{S}(x,z)\,k_{S}(z,x')\, \mathrm{d}z ,
\end{equation*}

\begin{equation*}
k_{S}(z,x')
=
\exp\left[
-\frac{1}{2}
\left(
\frac{d_{S}(z,x')}{\sigma_S}
\right)^{2}
\right].
\end{equation*}
We finally define 
\begin{equation*}
    G_S(x, x') = d_S(x, x') + \epsilon_g  h_{S}(x,x'),
\end{equation*}
where $d_S(x, x')$ denotes the pairwise geodesic distance on the surface $S$, and $\epsilon_g$ and $\sigma_S$ are parameters controlling the contribution of the global structure.
So our final structural/geometric term is defined as
\begin{equation*}
\begin{aligned}
D_{\text{Structure}}^{\text{soft}}(S_\alpha, S_\beta)
=\;
\eta\,\Bigg(
\int_{S_\alpha^2}\!\!\int_{S_\beta^2}\frac{1}{2}
\bigl| G_{\alpha}(x, x') - G_{\beta}(y, y') \bigr|^{q} \\
\, p(x,y) p(x',y') \, d\mu^\alpha(x)\, d\mu^\alpha(x')\,
d\mu^\beta(y)\, d\mu^\beta(y') \\+ (S_\alpha \leftrightarrow S_\beta) 
\Bigg)^{\frac{1}{q}}.
\end{aligned}
\end{equation*}\\

\nsb{The IFACE Distance.}
We now define our distance, referred to as the \textbf{IFACE
  distance}, using the soft-correspondence values, based on the top-$k$ normalized weights and
their associated vertices. For each point, we select the $k (=2)$ largest
values from the joint correspondence density weighted by local
geometric measures: 
\begin{equation*}
\begin{aligned}
\tilde{w}^{y_k}(x) &\equiv \text{Normalize}\left( \text{Top}^y_k \left[ p(x, y) \cdot \sqrt{|g_{\alpha}(x)| \cdot |g_{\beta}(y)|} \right] \right), \\
\tilde{w}^{x_k}(y) &\equiv \text{Normalize}\left( \text{Top}^x_k \left[ p(x, y) \cdot \sqrt{|g_{\alpha}(x)| \cdot |g_{\beta}(y)|} \right] \right).
\end{aligned}
\end{equation*}

Here, $p(x, y)$ denotes the soft correspondence \emph{density} between
points $x \in S_\alpha$ and $y \in S_\beta$, and $|g_{\alpha}(x)|$,
$|g_{\beta}(y)|$ are the determinants of the metric tensors at $x$ and
$y$, respectively. The \texttt{Top} operator selects the $k$ largest
weighted densities for each fixed variable (either $x$ or $y$), and
the \texttt{Normalize} operation ensures that the resulting weights
form a valid probability distribution, \ie, they
integrate to $1$ over the selected $k$-neighborhood. 

Now, $\tilde{w}^{y_k}(x)$ represents a normalized soft distribution over the top-$k$ points ($y_k$) on $S_\beta$ that most strongly correspond to a given point $x \in S_\alpha$, and vice versa for $\tilde{w}^{x_k}(y)$.

Then the IFACE distance is defined as soft distance defined by the new weights as 
\begin{equation}\label{eq: soft_global_distance_iface}
\begin{aligned}
D_{\text{IFACE}}(S_\alpha, S_\beta) = \sum_m\,\zeta_m \, D_\text{field}^m(S_\alpha, S_\beta) +  \eta \, D_\text{structure}(S_\alpha, S_\beta)~,
\end{aligned}
\end{equation}
where 
\begin{equation*}
\begin{aligned}
D_\text{field}^m(S_\alpha, S_\beta) = \frac{1}{2^{1/p}} \Bigg(\sum_k \int_{S_\alpha} | f_\alpha^m(x) - f_\beta^m(y_k) |^{p}\,  \tilde{w}^{y_{k}}(x)\, d^2x  \\
+ \int_{S_\beta} | f_\alpha^m(x_k) - f_\beta^m(y) |^{p}\,    \tilde{w}^{x_{k}}(y)\, d^2y \Bigg)^{\frac{1}{p}}
\end{aligned}
\end{equation*}
\begin{equation*}
\begin{aligned}
D_\text{structure}(S_\alpha, S_\beta) = \frac{1}{2^{1/q}}\sum_{k, k'}\Bigg(\int_{S_\alpha^2}\| G_{\alpha}(x, x') - G_{\beta}(y_k, y'_{k'}) \|^{q}\\ \tilde{w}^{y_{k}} (x) \, \tilde{w}^{y_{k'}}(x') \, d^2x \, d^2x'
+\\ \int_{S_\beta^2} \| G_{\alpha}(x_k, x^{'}_{k'}) - G_{\beta}(y, y') \|^{q}\, \tilde{w}^{x_{k}} (y) \, \tilde{w}^{x_{k'}}(y') \, d^2y \, d^2y' \Bigg)^{\frac{1}{q}}.
\end{aligned}
\end{equation*}

\section{\sb{Discrete Protein Surface}}\label{appendix: marginal distribution}
We use the discrete form of the problem as described in the main text. Here, we describe the protein surface generation pipeline,  calculation of the marginal distribution used in the method,  and the optimization process.\\

\nsb{Protein Surface Generation.}
We use a modified MaSIF pipeline (\href{https://github.com/jomimc/masif_molecule}{github.com/jomimc/masif\_molecule}) to generate protein surfaces comprising vertex positions, connectivity information, and feature files describing vertex-level attributes, including charge, hydrophobicity (Kyte--Doolittle), Gaussian curvature, mean curvature, amino acid identity, and the sequence index of the corresponding residue. The resulting protein meshes are preprocessed for analysis and contain a maximum of 3{,}000 vertices per mesh.

Because optimization becomes computationally expensive for large mesh sizes, we simplify protein surface meshes while consistently transferring per-vertex geometric and chemical properties to the reduced representation. Starting from an input surface mesh, we first perform a cleaning step to remove invalid elements and disconnected components, producing a watertight mesh. The cleaned mesh is then decimated using quadric edge collapse to obtain a target resolution, with the number of faces proportional to the desired vertex count. After decimation, small isolated components are removed, and the mesh is further refined using Laplacian smoothing to reduce discretization noise while preserving overall surface geometry.

To propagate vertex-wise properties from the original mesh to the simplified mesh, we establish a correspondence based on spatial proximity. Specifically, each vertex of the simplified mesh is mapped to its nearest vertex on the original mesh using a KD-tree search. For each mapped vertex, we define a local neighborhood on the original mesh using a fixed number of hops in the vertex adjacency graph. Continuous-valued properties are transferred by averaging over these neighborhoods, whereas discrete or categorical properties are directly inherited from the corresponding nearest vertices.

Finally, we compute geometric descriptors, including mean and Gaussian curvature, directly on the simplified mesh. The resulting meshes and associated properties are saved for downstream analysis, ensuring a consistent and computationally efficient surface representation with a bounded number of vertices.\\

\nsb{Calculation for marginal distributions.}
In this section, we provide the procedure used to calculate marginal distribution for the protein surfaces. Let $\{\mu_i^{\alpha}\}$ and $\{\mu_j^{\beta}\}$ denote the vertex areas of surfaces $S_\alpha$ and $S_\beta$, respectively. Let $f_i^{\alpha m}$ and $f_j^{\beta m}$ denote the feature fields, where
$m \in$ (Electrostatic Potential, Hydrophobicity, Hydrogen Bond Propensity, Mean curvature). We compute the marginal distributions $\rho_i^{\alpha}$ and $\rho_j^{\beta}$ as follows.
\paragraph{Step 1: Area-based density (constant field).}
\begin{equation*}
\rho_{i}^{\alpha (0)} = \frac{\mu_i^{\alpha}}{\sum_i \mu_i^{\alpha}},
\qquad
\rho_{j}^{S_\beta (0)} = \frac{\mu_j^{\beta}}{\sum_j \mu_j^{\beta}}.
\end{equation*}

\paragraph{Step 2: Global minimum feature value.}
\begin{equation*}
f_{\min}^{(m)} =
\min \Big( \min_i f_{i}^{\alpha (m)}, \; \min_j f_{j}^{\beta  (m)} \Big).
\end{equation*}

\paragraph{Step 3: Area-weighted, shifted features.}
\begin{equation*}
\tilde{f}_{i}^{\alpha (m)} = \big( f_{i}^{\alpha (m)} - f_{\min}^{(m)} \big)\, \mu_i^{\alpha},
\qquad
\tilde{f}_{j}^{\beta (m)} = \big( f_{j}^{\beta (m)} - f_{\min}^{(m)} \big)\, \mu_j^{\beta}.
\end{equation*}

\paragraph{Step 4: Feature normalization.}
\begin{equation*}
\tilde{f}_{i}^{\alpha (m)} \leftarrow
\frac{ \tilde{f}_{i}^{\alpha (m)} }{ \sum_i \tilde{f}_{i}^{\alpha (m)} },
\qquad
\tilde{f}_{j}^{\beta (m)} \leftarrow
\frac{ \tilde{f}_{j}^{\beta (m)} }{ \sum_j \tilde{f}_{j}^{\beta (m)} }.
\end{equation*}

\paragraph{Step 5: Combine area and feature distributions.}
\begin{equation*}
\rho_{i}^{\alpha} = \rho_{i}^{\alpha (0)} + \sum_m \tilde{f}_{i}^{\alpha (m)},
\qquad
\rho_{j}^{\beta} = \rho_{j}^{\beta (0)} + \sum_m \tilde{f}_{j}^{\beta (m)}.
\end{equation*}

\paragraph{Step 6: Final normalization.}
\begin{equation*}
\rho_i^{\alpha} \leftarrow \frac{\rho_i^{\alpha}}{\sum_i \rho_i^{\alpha}},
\qquad
\rho_j^{\beta} \leftarrow \frac{\rho_j^{\beta}}{\sum_j \rho_j^{\beta}}.
\end{equation*}\\

\nsb{Optimization for finding optimal coupling matrix.}\\

\noindent\emph{Optimization.---}
\begin{figure}[t]
\noindent
\fbox{%
  \begin{minipage}{0.95\linewidth}
  \textbf{Algorithm 1: Protein Mesh Alignment and Coupling Matrix Estimation}

  \medskip
\raggedright
\hspace*{4.5mm}\textbf{Input:} Surface meshes $S_\alpha$ and $S_\beta$ with vertex sets \\
\hspace*{14mm}$\{x_i\}, \{y_j\}$, associated vertex areas 
$\{\mu_i^\alpha\}, \{\mu_j^\beta\}$, \\\hspace*{14.5mm}and $m$-feature fields 
$f_i^{\alpha(m)}, f_j^{\beta(m)}$.
\medskip

 \raggedright
\hspace*{4.5mm}\textbf{Output:} Optimal coupling matrix $P_{ij}$.
  \begin{tabbing}
  000 \= 000 \= \kill
  1.\> Perform rigid alignment followed by \\ \hspace*{5.3mm}feature-aware nonrigid alignment.\\
  2.\> Transform vertices and features: \\
     \> $x_i \to \tilde{x}_i,\;
        f_i^{\alpha(m)} \to \tilde{f}_i^{\alpha(m)}$. \\
  3.\> Construct descriptor vectors \\
     \> $\tilde{X}_i = [\tilde{x}_i,\tilde{f}_i^{\alpha(m)}]$, 
        $Y_j = [y_j, f_j^{\beta(m)}]$. \\
  4.\> Compute Gaussian affinity kernel 
     $A_{ij} = \exp(-\|\tilde{X}_i-Y_j\|^2/\delta^2)$ \\\hspace*{5.5mm}with $\delta=0.01$. \\
  5.\> Apply Sinkhorn normalization to the Gaussian affinity matrix. \\
  6.\> Obtain initial coupling $P_{ij}$. \\
  7.\> Optimize Eq.~\eqref{eq:supp_loss_fn} with $\lambda=0.9$. \\
  8.\> Update $P_{ij}$. \\
  9.\> Re-optimize Eq.~\eqref{eq:supp_loss_fn} without entropic regularization. \\
  10.\> Derive final coupling matrix $P_{ij}$. \\
  11.\> Return $P_{ij}$. \\
  \end{tabbing}
  \end{minipage}
}
\Label{fig:algorithm1}
\end{figure}
We perform a four-step refinement to obtain the coupling matrix $P_{ij}$, which includes mesh alignment followed by optimization steps. We first perform rigid alignment of the meshes using RANSAC~\cite{fischler1981random} for coarse initialization, followed by Iterative Closest Point (ICP)~\cite{besl1992method, chen1992object}  for refinement. After that, we use non-rigid alignment in the combined real and feature space, employing the coherent point drift algorithm~\cite{myronenko2010point}. We convert the point sets into a Gaussian kernel to obtain a similarity measure between the transformed and target points. Then, we convert the Gaussian kernel into a coupling/soft-correspondence matrix using the Sinkhorn update~\cite{sinkhorn1967diagonal,  cuturi2013sinkhorn} with the marginal distribution. We feed that solution into the entropically regularized objective function for the field and structural terms with $\lambda = 0.9$ (this value is fixed across all examples). After that, we refine the coupling matrix by solving the objective function without entropic regularization.
The algorithmic steps are given in Algorithm 1.\\

\noindent\emph{Non-rigid alignment for initializing coupling matrix.---}
After rigid alignment of meshes, we perform non-rigid registration in the joint geometric and feature space using the Coherent Point Drift (CPD) algorithm~\cite{myronenko2010point}. CPD models the source points as centroids of a Gaussian mixture and seeks a smooth non-rigid transformation that maximizes the likelihood of the target points under this model while enforcing motion coherence. Let $X \equiv [x_i, f_i^{\alpha (m)}]$ and $Y \equiv [y_j, f_j^{\beta (m)}]$ denote the source and target vertex sets, respectively, after rigid alignment in coordinate space, together with the $m$-feature-field values defined on those vertices.
The non-rigid CPD transformation $T(\cdot)$ produces transformed points for source $\tilde{X}=\{T(X_i)\}$. 

To quantify the similarity between transformed and target points, we construct a Gaussian kernel
\begin{equation*}
A_{ij}
=
\exp\!\left(
-\frac{\|\tilde{X}_i-Y_j\|^2}{\delta^2}
\right),
\end{equation*}
which encodes pairwise affinities between $\tilde{X}$ and $Y$. We then interpret $A$ as an entropic-regularized similarity matrix and convert it into a soft correspondence (coupling) matrix by applying Sinkhorn iterations~\cite{sinkhorn1967diagonal,  cuturi2013sinkhorn} to match prescribed marginal distributions. This yields a coupling matrix $P_{i\,j}$, which serves as the initial coupling (soft correspondence) matrix. We set $\delta = 0.01$ to enforce a tight similarity constraint. All steps are performed on unit-normalized features. For each feature field, joint min--max normalization is applied, where the scaling parameters are computed from the concatenation of the corresponding features and the values are mapped to $[-1, 1]$. Meshes are independently normalized to lie within the unit sphere.\\

\noindent\emph{Entropic and entropic free regularized optimization of objective function.---}
We use the initial coupling matrix $P_{i\, j}$ obtained using the above procedure as an initialization for the entropic optimization of the objective function combining the structural and field terms:
\begin{equation*}
\begin{split}
P_{ij}^{entro} = \arg \min_{P_{i\,j}} \Bigl[
(1 - \lambda)\,\mathcal{F}(S_\alpha, S_\beta)
+ \lambda\,\mathcal{S}(S_\alpha, S_\beta) \\
{} + \epsilon\, P_{ij}\log P_{ij}
\Bigr],
\end{split}
\end{equation*}
\begin{equation}
\label{eq:supp_loss_fn}
\text{subject to} \,
\sum_j P_{ij} = \rho_{i}^{\alpha}, \quad
\sum_i P_{ij} = \rho_{j}^{\beta},
\end{equation}
where $\mathcal{F}(S_\alpha, S_\beta)$ and $\mathcal{S}(S_\alpha, S_\beta)$ are the field and structural terms defined in the Methods section of the main text for the optimization.We use unit-normalized meshes to compute the distance matrices used in the structural term.

\noindent\emph{Weight parameters.---}
The parameter $\lambda$ controls the balance between geometric correspondence and field alignment in the surface coupling matrix. We set $\lambda = 0.9$, giving dominant weight to the structural term so that the surface map remains geometrically continuous. The field term then acts as a secondary constraint that aligns regions with similar physicochemical fields. While this contribution can break geometric symmetries, the strong structural weighting preserves the global organization of the correspondence.

To verify that this choice achieves the intended balance, we examined a controlled toy system consisting of ellipsoidal surfaces. The surfaces were geometrically identical but carried different charge distributions, allowing the influence of the field term to be isolated from geometry. Across the explored range of $\lambda$, smaller values emphasize agreement of the field term, whereas larger values emphasize geometric consistency. We observe qualitatively stable behavior for $\lambda \gtrsim 0.85$ and therefore select $\lambda = 0.9$, which maintains structural continuity while allowing the field contribution to guide feature-consistent alignment across the surface.

\noindent\emph{Entropic regularization parameter.---}
The entropic regularization parameter $\epsilon$ is chosen adaptively to maintain numerical stability during optimization. Specifically,
\begin{equation*}
\epsilon = \max\left(
\frac{\max \mathcal{C}}{\kappa},
\; \tau \cdot \mathrm{median}(\mathcal{C}),
\; 10^{-6}
\right),
\end{equation*}
where $\mathcal{C}$ denotes the collection of cost matrices used in the optimization, including the structural matrices $G_{\alpha}$, $G_{\beta}$ and the 
field-based cost term comparing aligned physicochemical features
$\mathcal{F}(S_{\alpha}, S_{\beta})$. 
The geometric and chemical terms are scaled to comparable magnitudes, so that $\lambda$ controls their relative contribution rather than compensating for differences in scale.
The maximum and median are taken over all scalar entries of these matrices. The constant $\kappa$ prevents numerical underflow, while $\tau$ sets the regularization scale relative to the typical magnitude of the cost terms; we use $\kappa = 700$ and $\tau = 0.05$. This choice was found to be robust across datasets and does not affect the qualitative behavior of the distances.

The entropically regularized problem is first solved to obtain an initial coupling $P_{ij}^{\mathrm{entro}}$. This solution is then used to initialize a second optimization without entropic regularization ($\epsilon = 0$), yielding the final optimal coupling matrix $P_{ij}$. All optimizations were performed using the POT Python library~\cite{flamary2021pot,flamary2024pot}. Across the dataset of 1080 protein pairs, the full pipeline---including optimization and distance computations---takes about 8.5 minutes per pair on average. Experiments were conducted on a workstation equipped with an Intel Core i9-14900K CPU (24 cores, 32 threads) and 64~GB RAM running Ubuntu~24.04.3~LTS. For rapid screening and large-scale comparison, meshes containing approximately 1,000 vertices substantially reduce the computational cost while preserving comparable classification and clustering performance in the present benchmarks. The feature-specific scaling relations described in Sec. (Normalization scale for distances) were used to transfer the fixed normalization ranges between mesh resolutions.\\

\nsb{Color transfer between the mapped surfaces.}
Colors on the target surface $c^{\mathrm{target}}_j$  were computed by aggregating probability-weighted colors from matched vertices on the source surface $c^{\mathrm{source}}_i$ (e.g., target: \texttt{6XDS} and source: \texttt{6XRX} in Fig. \ref{fig:combined-protein}), followed by normalization by the total incoming probability mass of the coupling matrix:
$$
c^{\mathrm{target}}_j =
\frac{\sum_i P_{i \rightarrow j}\, c^{\mathrm{source}}_i}
{\sum_i P_{i \rightarrow j}}.
$$
Here, the summation is restricted to the top two matches per $i$.\\

\section{\sb{Comments on Geometry-Based Correspondence Methods}}

We assessed geometry-driven correspondence methods on protein surfaces using a representative functional correspondence approach~\cite{ovsjanikov2012functional}, widely used in shape matching, as a benchmark. These methods did not consistently distinguish structural divergence from conformational variability on the dataset considered in the main text. For brevity, we do not present quantitative benchmarks here.
\\
These observations suggest that the assumptions underlying such methods—particularly the reliance on near-isometric structure for stable spectral representations—may be frequently violated for protein surfaces in a functionally meaningful way. Furthermore, the geometric descriptors (e.g., Heat Kernel Signature~\cite{sun2009concise} and Wave Kernel Signature~\cite{aubry2011wave}) employed in these approaches may be insufficient to capture the high geometric complexity and roughness characteristic of protein surfaces, thereby limiting their effectiveness in this setting.
\\
Taken together, these observations highlight the limitations of purely geometric approaches and motivate the joint geometric--chemical formulation introduced in IFACE.
\\

\section{\sb{Normalization scale for distances}}
\Label{sec:si_normalization}

To avoid recomputing the normalization ranges from each evaluation
dataset, we constructed a fixed reference set containing 130 protein
surfaces. Of these, 65 were selected from the test partition of the
SHREC 2021 protein-surface benchmark by choosing one representative
surface from each of its 65 PDB-based classes
\cite{SHREC2021dataset}. In the SHREC PDB-based classification, each
class contains the available NMR conformations associated with the
same PDB entry. Selecting one surface per class therefore avoids
overrepresenting proteins for which a larger number of conformations
is available. The remaining 65 surfaces were drawn from the family
and conformer collections considered in this study.

The complete reference set generates $8,385$
distinct unordered, non-self protein-pair comparisons. Because these pairs span proteins drawn from several structural classes and datasets, they produce broad distributions of structural and physicochemical
distances. The resulting extrema therefore provide a broad empirical normalization scale for the present analysis. These extrema should, however, be interpreted as empirical reference bounds rather than absolute theoretical bounds on the corresponding distances.

All reference surfaces were sampled at $N_0=1000$ vertices to reduce the computational cost of evaluating the 8,385 pairwise comparisons. For each distance component $f$, the global minimum and maximum were computed over the reference set.

The reference ranges were transferred to the 3000-vertex benchmark resolution using the descriptor-specific scaling relation
\begin{equation*}
d_f(N)=
d_f(N_0)
\left(\frac{N}{N_0}\right)^{p_f},
\Label{eq:distance_scaling}
\end{equation*}
where $p_f$ is the scaling exponent for component $f$. For each of the $K=1080$ protein pairs evaluated at both resolutions,
\begin{equation*}
p_{f,k}=
\frac{
\log\left[d_{f,k}(3000)/d_{f,k}(1000)\right]
}{
\log(3000/1000)
},
\end{equation*}
and the final exponent was obtained as
\begin{equation*}
p_f=\frac{1}{K}\sum_{k=1}^{K}p_{f,k}.
\end{equation*}

The fitted exponents were used to scale the reference extrema from 1000 to 3000 vertices,
\begin{equation*}
d_{f,\min/\max}(3000)=
d_{f,\min/\max}(1000)\,3^{p_f}.
\end{equation*}

\begin{table}[t]
\centering
\caption{Descriptor-specific approximate scaling exponents and fixed normalization ranges transferred from 1,000 to 3,000 vertices.}
\Label{tab:reference_normalization_ranges}
\begin{tabular}{lrrr}
\toprule
\textbf{Distance}
& $\boldsymbol{p_f}$
& \textbf{Reference minimum }
& \textbf{Reference maximum} \\
\midrule
Structural
& 0.9 & 3398 & 104856 \\

Mean curvature
& 1.3 & 490 & 2038 \\

Hydrogen bond
& 1.3 & 225 & 2008 \\

Hydrophobicity
& 1.3 & 3367 & 35990 \\
 
Electrostatics
& 1.2 & 1220 & 125866 \\
\bottomrule
\end{tabular}
\end{table}

Each raw distance was then scaled by the corresponding reference range,
\begin{equation*}
\bar{d}_f=
\frac{d_f}
{d_{f,\max}-d_{f,\min}}.
\Label{eq:fixed_range_normalization}
\end{equation*}
The reference minimum is not subtracted such that zero raw distance remains zero.
The same fixed ranges were used for all family and conformer benchmarks, avoiding benchmark-specific rescaling while allowing the normalization scale to be transferred across mesh resolutions.

\section{\sb{Full Ablation Study}}
\begin{table*}[t]
\centering
\caption{Ablation study of correspondence-based and distribution-based (JSD) feature combinations for binary classification of inter- and intra-protein protein pairs for conformer benchmark. For each correspondence-based feature combination, the structural distance is paired with its distribution-based counterpart, the JSD Gaussian curvature distance.}
\label{tab:ablation_compare_conf}
\scriptsize
\setlength{\tabcolsep}{4pt}
\begin{tabular}{lcc|cc}
\toprule
\multirow{2}{*}{Feature Combination} & \multicolumn{2}{c|}{Correspondence-based} & \multicolumn{2}{c}{Distribution-based (JSD)}\\
\cmidrule(r){2-3}\cmidrule(l){4-5}
& AP & ROC-AUC & AP & ROC-AUC\\
\midrule
Electrostatic & 0.9736 & 0.9879 & 0.9501 & 0.9847\\
HBond & 0.9226 & 0.9581 & 0.3894 & 0.7018\\
Hydrophobicity & 0.9013 & 0.9513 & 0.7314 & 0.8996\\
Mean Curvature & 0.8107 & 0.8926 & 0.3299 & 0.6192\\
Structural & 0.8939 & 0.9166 & 0.2658 & 0.5745\\
\midrule
Electrostatic + Hydrophobicity & 0.9452 & 0.9734 & 0.9933 & 0.9977\\
HBond + Electrostatic & 0.9673 & 0.9817 & 0.9533 & 0.9861\\
HBond + Hydrophobicity & 0.9215 & 0.9584 & 0.7629 & 0.9072\\
Mean Curvature + Electrostatic & 0.8618 & 0.9298 & 0.9302 & 0.9784\\
Mean Curvature + HBond & 0.8568 & 0.9246 & 0.4029 & 0.7009\\
Mean Curvature + Hydrophobicity & 0.8822 & 0.9390 & 0.7108 & 0.8820\\
Structural + Electrostatic & 0.9111 & 0.9274 & 0.9386 & 0.9821\\
Structural + HBond & 0.9138 & 0.9274 & 0.4007 & 0.7007\\
Structural + Hydrophobicity & 0.9205 & 0.9349 & 0.7307 & 0.8875\\
Structural + Mean Curvature & 0.9118 & 0.9281 & 0.3210 & 0.6299\\
\midrule
HBond + Electrostatic + Hydrophobicity & 0.9490 & 0.9737 & 0.9929 & 0.9975\\
Mean Curvature + Electrostatic + Hydrophobicity & 0.9111 & 0.9550 & 0.9868 & 0.9951\\
Mean Curvature + HBond + Electrostatic & 0.8928 & 0.9468 & 0.9270 & 0.9787\\
Mean Curvature + HBond + Hydrophobicity & 0.8970 & 0.9467 & 0.7252 & 0.8862\\
Structural + Electrostatic + Hydrophobicity & 0.9265 & 0.9418 & 0.9923 & 0.9974\\
Structural + HBond + Electrostatic & 0.9221 & 0.9355 & 0.9470 & 0.9841\\
Structural + HBond + Hydrophobicity & 0.9240 & 0.9409 & 0.7579 & 0.8964\\
Structural + Mean Curvature + Electrostatic & 0.9209 & 0.9363 & 0.9185 & 0.9756\\
Structural + Mean Curvature + HBond & 0.9186 & 0.9350 & 0.4102 & 0.7026\\
Structural + Mean Curvature + Hydrophobicity & 0.9205 & 0.9402 & 0.6997 & 0.8692\\
\midrule
Mean Curvature + HBond + Electrostatic + Hydrophobicity & 0.9202 & 0.9592 & 0.9863 & 0.9949\\
Structural + HBond + Electrostatic + Hydrophobicity & 0.9287 & 0.9467 & 0.9919 & 0.9972\\
Structural + Mean Curvature + Electrostatic + Hydrophobicity & 0.9269 & 0.9463 & 0.9848 & 0.9946\\
Structural + Mean Curvature + HBond + Electrostatic & 0.9255 & 0.9424 & 0.9202 & 0.9763\\
Structural + Mean Curvature + HBond + Hydrophobicity & 0.9231 & 0.9444 & 0.7182 & 0.8762\\
\midrule
All Features & 0.9284 & 0.9496 & 0.9837 & 0.9942\\
\bottomrule
\end{tabular}
\end{table*}

\begin{table*}[t]
\centering
\caption{Ablation study of correspondence-based and distribution-based (JSD) feature combinations for binary classification of inter- and intra-family protein pairs. For each correspondence-based feature combination, the structural distance is paired with its distribution-based counterpart, the JSD Gaussian curvature distance.}
\label{tab:ablation_compare}
\scriptsize
\setlength{\tabcolsep}{4pt}
\begin{tabular}{lcc|cc}
\toprule
\multirow{2}{*}{Feature Combination} & \multicolumn{2}{c|}{Correspondence-based} & \multicolumn{2}{c}{Distribution-based (JSD)}\\
\cmidrule(r){2-3}\cmidrule(l){4-5}
& AP & ROC-AUC & AP & ROC-AUC\\
\midrule
Electrostatic & 0.4650 & 0.7411 & 0.3582 & 0.7128\\
HBond & 0.6402 & 0.8068 & 0.3556 & 0.7951\\
Hydrophobicity & 0.5874 & 0.8617 & 0.5019 & 0.8277\\
Mean Curvature & 0.8013 & 0.8985 & 0.7003 & 0.9336\\
Structural & 0.4787 & 0.8371 & 0.6134 & 0.9056\\
\midrule
Electrostatic + Hydrophobicity & 0.5919 & 0.8659 & 0.4720 & 0.8131\\
HBond + Electrostatic & 0.6535 & 0.8107 & 0.4591 & 0.7950\\
HBond + Hydrophobicity & 0.6252 & 0.8359 & 0.5007 & 0.8323\\
Mean Curvature + Electrostatic & 0.8058 & 0.9061 & 0.4688 & 0.8004\\
Mean Curvature + HBond & 0.8108 & 0.8748 & 0.7321 & 0.9038\\
Mean Curvature + Hydrophobicity & 0.8277 & 0.9215 & 0.7665 & 0.8990\\
Structural + Electrostatic & 0.5023 & 0.8395 & 0.4659 & 0.8004\\
Structural + HBond & 0.7290 & 0.9342 & 0.7787 & 0.9162\\
Structural + Hydrophobicity & 0.7197 & 0.9260 & 0.7977 & 0.9024\\
Structural + Mean Curvature & 0.7313 & 0.9514 & 0.6981 & 0.9320\\
\midrule
HBond + Electrostatic + Hydrophobicity & 0.6279 & 0.8399 & 0.4961 & 0.8266\\
Mean Curvature + Electrostatic + Hydrophobicity & 0.8301 & 0.9261 & 0.5696 & 0.8483\\
Mean Curvature + HBond + Electrostatic & 0.8168 & 0.8803 & 0.5613 & 0.8357\\
Mean Curvature + HBond + Hydrophobicity & 0.8042 & 0.8983 & 0.7209 & 0.8905\\
Structural + Electrostatic + Hydrophobicity & 0.7157 & 0.9247 & 0.5938 & 0.8553\\
Structural + HBond + Electrostatic & 0.7210 & 0.9318 & 0.5796 & 0.8432\\
Structural + HBond + Hydrophobicity & 0.8469 & 0.9562 & 0.7697 & 0.9014\\
Structural + Mean Curvature + Electrostatic & 0.7316 & 0.9520 & 0.5190 & 0.8481\\
Structural + Mean Curvature + HBond & 0.9434 & 0.9838 & 0.8241 & 0.9299\\
Structural + Mean Curvature + Hydrophobicity & 0.8556 & 0.9736 & 0.8526 & 0.9241\\
\midrule
Mean Curvature + HBond + Electrostatic + Hydrophobicity & 0.8070 & 0.9014 & 0.5787 & 0.8560\\
Structural + HBond + Electrostatic + Hydrophobicity & 0.8397 & 0.9543 & 0.6138 & 0.8646\\
Structural + Mean Curvature + Electrostatic + Hydrophobicity & 0.8523 & 0.9730 & 0.6606 & 0.8792\\
Structural + Mean Curvature + HBond + Electrostatic & 0.9401 & 0.9826 & 0.6335 & 0.8687\\
Structural + Mean Curvature + HBond + Hydrophobicity & 0.9513 & 0.9835 & 0.8429 & 0.9231\\
\midrule
All Features & 0.9487 & 0.9829 & 0.6799 & 0.8866\\
\bottomrule
\end{tabular}
\end{table*}

\begin{table*}[t]
\centering
\caption{Adjusted Rand Index (ARI) for family-level clustering using all clustering algorithms and distance measures. Higher values indicate better agreement between predicted clusters and the ground-truth protein family labels.}
\label{tab:clustering_ari_all}
\scriptsize
\setlength{\tabcolsep}{3pt}
\begin{tabular}{lccccccc}
\toprule
\textbf{Algorithm} & \textbf{IFACE} & \textbf{JSD All} & \textbf{LB} & \textbf{MaSIF} & \textbf{MaSIF} & \textbf{SurfaceID} & \textbf{SurfaceID} \\
 & \textbf{(All Features)} & \textbf{Features} & \textbf{Distance} & \textbf{Mean} & \textbf{Mean+Max} & \textbf{Mean} & \textbf{Mean+Max} \\
\midrule
Single & 0.653 & 0.300 & 0.188 & 0.537 & 0.417 & 0.205 & 0.153\\
Complete & 0.899 & 0.556 & 0.424 & 0.578 & 0.380 & 0.218 & 0.328\\
Average & 0.889 & 0.358 & 0.473 & 0.631 & 0.408 & 0.249 & 0.114\\
Weighted & 0.889 & 0.358 & 0.513 & 0.578 & 0.445 & 0.262 & 0.328\\
\midrule
CMDS(4D)+K-means & 0.739 & 0.476 & 0.512 & 0.681 & 0.566 & 0.358 & 0.328\\
CMDS(4D)+Ward & 0.739 & 0.451 & 0.512 & 0.713 & 0.558 & 0.249 & 0.439\\
CMDS(5D)+K-means & 0.899 & 0.571 & 0.580 & 0.631 & 0.697 & 0.358 & 0.399\\
CMDS(5D)+Ward & 0.899 & 0.571 & 0.437 & 0.631 & 0.697 & 0.249 & 0.281\\
CMDS(6D)+K-means & 0.899 & 0.565 & 0.512 & 0.578 & 0.697 & 0.358 & 0.255\\
CMDS(6D)+Ward & 0.899 & 0.495 & 0.437 & 0.578 & 0.697 & 0.249 & 0.242\\
CMDS(8D)+K-means & 1.000 & 0.392 & 0.682 & 0.631 & 0.776 & 0.358 & 0.308\\
CMDS(8D)+Ward & 1.000 & 0.357 & 0.748 & 0.631 & 0.697 & 0.249 & 0.328\\
CMDS(10D)+K-means & 0.936 & 0.502 & 0.812 & 0.631 & 0.697 & 0.358 & 0.190\\
CMDS(10D)+Ward & 0.949 & 0.502 & 0.748 & 0.631 & 0.566 & 0.249 & 0.328\\
CMDS(15D)+K-means & 0.899 & 0.357 & 0.682 & 0.631 & 0.566 & 0.358 & 0.255\\
CMDS(15D)+Ward & 0.899 & 0.502 & 0.748 & 0.631 & 0.493 & 0.249 & 0.328\\
CMDS(20D)+K-means & 0.923 & 0.412 & 0.512 & 0.631 & 0.512 & 0.358 & 0.255\\
CMDS(20D)+Ward & 0.899 & 0.357 & 0.748 & 0.631 & 0.697 & 0.249 & 0.328\\
CMDS(80\%)+K-means & 0.655 & 0.309 & 0.369 & 0.586 & 0.566 & 0.164 & 0.328\\
CMDS(80\%)+Ward & 0.936 & 0.309 & 0.412 & 0.586 & 0.697 & 0.127 & 0.439\\
CMDS(90\%)+K-means & 0.899 & 0.392 & 0.512 & 0.627 & 0.512 & 0.164 & 0.255\\
CMDS(90\%)+Ward & 0.923 & 0.357 & 0.512 & 0.618 & 0.697 & 0.127 & 0.242\\
CMDS(95\%)+K-means & 0.655 & 0.357 & 0.512 & 0.631 & 0.512 & 0.358 & 0.308\\
CMDS(95\%)+Ward & 0.899 & 0.502 & 0.437 & 0.631 & 0.493 & 0.249 & 0.328\\
CMDS(99\%)+K-means & 0.691 & 0.369 & 0.682 & 0.631 & 0.439 & 0.358 & 0.255\\
CMDS(99\%)+Ward & 0.899 & 0.357 & 0.748 & 0.631 & 0.697 & 0.249 & 0.328\\
\bottomrule
\end{tabular}
\end{table*}

\begin{table*}[t]
\centering
\caption{Cluster purity obtained using all clustering algorithms for each distance measure on the family benchmark. Higher values indicate better agreement between predicted clusters and the ground-truth protein family labels.}
\label{tab:clustering_purity_all}
\scriptsize
\setlength{\tabcolsep}{3pt}
\begin{tabular}{lccccccc}
\toprule
\textbf{Algorithm} & \textbf{IFACE} & \textbf{JSD All} & \textbf{LB} & \textbf{MaSIF} & \textbf{MaSIF} & \textbf{SurfaceID} & \textbf{SurfaceID} \\
 & \textbf{(All Features)} & \textbf{Features} & \textbf{Distance} & \textbf{Mean} & \textbf{Mean+Max} & \textbf{Mean} & \textbf{Mean+Max} \\
\midrule
Single & 0.800 & 0.640 & 0.600 & 0.840 & 0.680 & 0.600 & 0.560\\
Complete & 0.920 & 0.720 & 0.760 & 0.840 & 0.760 & 0.640 & 0.640\\
Average & 0.920 & 0.680 & 0.760 & 0.840 & 0.720 & 0.680 & 0.560\\
Weighted & 0.920 & 0.680 & 0.760 & 0.840 & 0.720 & 0.680 & 0.640\\
\midrule
CMDS(4D)+K-means & 0.920 & 0.720 & 0.800 & 0.880 & 0.800 & 0.720 & 0.640\\
CMDS(4D)+Ward & 0.920 & 0.720 & 0.800 & 0.920 & 0.800 & 0.680 & 0.720\\
CMDS(5D)+K-means & 0.920 & 0.760 & 0.840 & 0.840 & 0.840 & 0.720 & 0.640\\
CMDS(5D)+Ward & 0.920 & 0.760 & 0.760 & 0.840 & 0.840 & 0.680 & 0.680\\
CMDS(6D)+K-means & 0.920 & 0.720 & 0.800 & 0.840 & 0.840 & 0.720 & 0.640\\
CMDS(6D)+Ward & 0.920 & 0.680 & 0.760 & 0.840 & 0.840 & 0.680 & 0.640\\
CMDS(8D)+K-means & 1.000 & 0.720 & 0.880 & 0.840 & 0.880 & 0.720 & 0.640\\
CMDS(8D)+Ward & 1.000 & 0.680 & 0.880 & 0.840 & 0.840 & 0.680 & 0.640\\
CMDS(10D)+K-means & 0.960 & 0.800 & 0.920 & 0.840 & 0.840 & 0.720 & 0.600\\
CMDS(10D)+Ward & 0.960 & 0.800 & 0.880 & 0.840 & 0.800 & 0.680 & 0.640\\
CMDS(15D)+K-means & 0.920 & 0.680 & 0.880 & 0.840 & 0.800 & 0.720 & 0.640\\
CMDS(15D)+Ward & 0.920 & 0.800 & 0.880 & 0.840 & 0.800 & 0.680 & 0.640\\
CMDS(20D)+K-means & 0.960 & 0.720 & 0.800 & 0.840 & 0.760 & 0.720 & 0.640\\
CMDS(20D)+Ward & 0.920 & 0.680 & 0.880 & 0.840 & 0.840 & 0.680 & 0.640\\
CMDS(80\%)+K-means & 0.920 & 0.680 & 0.720 & 0.840 & 0.800 & 0.600 & 0.640\\
CMDS(80\%)+Ward & 0.960 & 0.680 & 0.760 & 0.840 & 0.840 & 0.560 & 0.720\\
CMDS(90\%)+K-means & 0.920 & 0.720 & 0.800 & 0.840 & 0.760 & 0.600 & 0.640\\
CMDS(90\%)+Ward & 0.960 & 0.680 & 0.800 & 0.840 & 0.840 & 0.560 & 0.640\\
CMDS(95\%)+K-means & 0.920 & 0.680 & 0.800 & 0.840 & 0.760 & 0.720 & 0.640\\
CMDS(95\%)+Ward & 0.920 & 0.800 & 0.760 & 0.840 & 0.800 & 0.680 & 0.640\\
CMDS(99\%)+K-means & 0.920 & 0.680 & 0.880 & 0.840 & 0.760 & 0.720 & 0.640\\
CMDS(99\%)+Ward & 0.920 & 0.680 & 0.880 & 0.840 & 0.840 & 0.680 & 0.640\\
\bottomrule
\end{tabular}
\end{table*}

\begin{table*}[t]
\centering
\caption{Leave-one-feature-out clustering ablation for IFACE on the family benchmark. Entries report Adjusted Rand Index (ARI) across hierarchical clustering and CMDS-based clustering variants.}
\label{tab:iface_clustering_loo_ari}
\scriptsize
\setlength{\tabcolsep}{3pt}
\renewcommand{\arraystretch}{1.08}
\begin{tabular}{lcccccccc}
\toprule
\textbf{Algorithm} & \textbf{All Features} & \textbf{w/o Structural} & \textbf{w/o Mean Curvature} & \textbf{w/o Hydrogen Bond} & \textbf{w/o Electrostatics} & \textbf{w/o Hydrophobicity} & \textbf{Chemical Only} & \textbf{Geometry Only} \\
\midrule
Single & 0.653 & 0.327 & 0.327 & 0.867 & 0.889 & 0.889 & 0.256 & 0.669 \\
Complete & 0.899 & 0.726 & 0.824 & 0.867 & 0.899 & 0.936 & 0.264 & 0.867 \\
Average & 0.889 & 0.388 & 0.699 & 0.867 & 0.889 & 0.899 & 0.388 & 0.867 \\
Weighted & 0.889 & 0.559 & 0.699 & 0.867 & 0.889 & 0.936 & 0.323 & 0.867 \\
\midrule
CMDS(4D)+K-means & 0.739 & 0.766 & 0.675 & 0.739 & 0.739 & 0.675 & 0.264 & 0.675 \\
CMDS(4D)+Ward & 0.739 & 0.756 & 0.675 & 0.739 & 0.739 & 0.675 & 0.213 & 0.675 \\
CMDS(5D)+K-means & 0.899 & 0.698 & 0.777 & 0.691 & 0.645 & 0.675 & 0.328 & 0.675 \\
CMDS(5D)+Ward & 0.899 & 0.698 & 0.840 & 0.691 & 0.899 & 0.691 & 0.540 & 0.691 \\
CMDS(6D)+K-means & 0.899 & 0.791 & 0.521 & 0.691 & 0.899 & 0.899 & 0.328 & 0.691 \\
CMDS(6D)+Ward & 0.899 & 0.791 & 0.485 & 0.691 & 0.899 & 0.899 & 0.540 & 0.691 \\
CMDS(8D)+K-means & 1.000 & 0.949 & 0.651 & 0.691 & 1.000 & 0.936 & 0.264 & 0.691 \\
CMDS(8D)+Ward & 1.000 & 0.911 & 0.912 & 0.660 & 1.000 & 0.899 & 0.540 & 0.691 \\
CMDS(10D)+K-means & 0.936 & 0.924 & 0.854 & 0.739 & 0.936 & 0.936 & 0.232 & 0.582 \\
CMDS(10D)+Ward & 0.949 & 0.849 & 0.854 & 0.739 & 0.949 & 0.936 & 0.540 & 0.691 \\
CMDS(15D)+K-means & 0.899 & 0.572 & 0.593 & 0.673 & 0.899 & 0.899 & 0.264 & 0.582 \\
CMDS(15D)+Ward & 0.899 & 0.923 & 0.811 & 0.691 & 0.923 & 0.936 & 0.559 & 0.691 \\
CMDS(20D)+K-means & 0.923 & 0.513 & 0.730 & 0.675 & 0.923 & 0.899 & 0.414 & 0.691 \\
CMDS(20D)+Ward & 0.899 & 0.923 & 0.854 & 0.675 & 1.000 & 0.936 & 0.328 & 0.675 \\
\midrule
CMDS(80\%)+K-means & 0.655 & 0.584 & 0.854 & 0.675 & 0.936 & 0.936 & 0.362 & 0.739 \\
CMDS(80\%)+Ward & 0.936 & 0.605 & 0.854 & 0.675 & 1.000 & 0.936 & 0.421 & 0.867 \\
CMDS(90\%)+K-means & 0.899 & 0.572 & 0.924 & 0.673 & 0.899 & 0.899 & 0.264 & 0.582 \\
CMDS(90\%)+Ward & 0.923 & 0.846 & 0.811 & 0.691 & 0.923 & 0.936 & 0.559 & 0.691 \\
CMDS(95\%)+K-means & 0.655 & 0.539 & 0.851 & 0.691 & 0.897 & 0.899 & 0.362 & 0.691 \\
CMDS(95\%)+Ward & 0.899 & 0.923 & 0.811 & 0.675 & 1.000 & 0.936 & 0.328 & 0.691 \\
CMDS(99\%)+K-means & 0.691 & 0.599 & 0.730 & 0.691 & 0.923 & 0.899 & 0.264 & 0.691 \\
CMDS(99\%)+Ward & 0.899 & 0.605 & 0.854 & 0.675 & 1.000 & 0.936 & 0.328 & 0.675 \\
\bottomrule
\end{tabular}
\end{table*}

\begin{table*}[t]
\centering
\caption{Leave-one-feature-out clustering ablation for IFACE on the family benchmark. Entries report cluster purity across hierarchical clustering and CMDS-based clustering variants.}
\label{tab:iface_clustering_loo_purity}
\scriptsize
\setlength{\tabcolsep}{3pt}
\renewcommand{\arraystretch}{1.08}
\begin{tabular}{lcccccccc}
\toprule
\textbf{Algorithm} & \textbf{All Features} & \textbf{w/o Structural} & \textbf{w/o Mean Curvature} & \textbf{w/o Hydrogen Bond} & \textbf{w/o Electrostatics} & \textbf{w/o Hydrophobicity} & \textbf{Chemical Only} & \textbf{Geometry Only} \\
\midrule
Single & 0.800 & 0.680 & 0.680 & 0.920 & 0.920 & 0.920 & 0.640 & 0.800 \\
Complete & 0.920 & 0.880 & 0.880 & 0.920 & 0.920 & 0.960 & 0.680 & 0.920 \\
Average & 0.920 & 0.680 & 0.800 & 0.920 & 0.920 & 0.920 & 0.680 & 0.920 \\
Weighted & 0.920 & 0.760 & 0.800 & 0.920 & 0.920 & 0.960 & 0.720 & 0.920 \\
\midrule
CMDS(4D)+K-means & 0.920 & 0.880 & 0.920 & 0.920 & 0.920 & 0.920 & 0.640 & 0.920 \\
CMDS(4D)+Ward & 0.920 & 0.880 & 0.920 & 0.920 & 0.920 & 0.920 & 0.640 & 0.920 \\
CMDS(5D)+K-means & 0.920 & 0.880 & 0.840 & 0.920 & 0.880 & 0.920 & 0.680 & 0.920 \\
CMDS(5D)+Ward & 0.920 & 0.880 & 0.880 & 0.920 & 0.920 & 0.920 & 0.760 & 0.920 \\
CMDS(6D)+K-means & 0.920 & 0.920 & 0.800 & 0.920 & 0.920 & 0.920 & 0.680 & 0.920 \\
CMDS(6D)+Ward & 0.920 & 0.920 & 0.760 & 0.920 & 0.920 & 0.920 & 0.760 & 0.920 \\
CMDS(8D)+K-means & 1.000 & 0.960 & 0.920 & 0.920 & 1.000 & 0.960 & 0.640 & 0.920 \\
CMDS(8D)+Ward & 1.000 & 0.960 & 0.960 & 0.880 & 1.000 & 0.920 & 0.760 & 0.920 \\
CMDS(10D)+K-means & 0.960 & 0.960 & 0.920 & 0.920 & 0.960 & 0.960 & 0.640 & 0.880 \\
CMDS(10D)+Ward & 0.960 & 0.920 & 0.920 & 0.920 & 0.960 & 0.960 & 0.760 & 0.920 \\
CMDS(15D)+K-means & 0.920 & 0.760 & 0.880 & 0.920 & 0.920 & 0.920 & 0.640 & 0.880 \\
CMDS(15D)+Ward & 0.920 & 0.960 & 0.880 & 0.920 & 0.960 & 0.960 & 0.760 & 0.920 \\
CMDS(20D)+K-means & 0.960 & 0.760 & 0.840 & 0.920 & 0.960 & 0.920 & 0.680 & 0.920 \\
CMDS(20D)+Ward & 0.920 & 0.960 & 0.920 & 0.920 & 1.000 & 0.960 & 0.680 & 0.920 \\
\midrule
CMDS(80\%)+K-means & 0.920 & 0.840 & 0.920 & 0.920 & 0.960 & 0.960 & 0.720 & 0.920 \\
CMDS(80\%)+Ward & 0.960 & 0.800 & 0.920 & 0.920 & 1.000 & 0.960 & 0.720 & 0.920 \\
CMDS(90\%)+K-means & 0.920 & 0.760 & 0.960 & 0.920 & 0.920 & 0.920 & 0.640 & 0.880 \\
CMDS(90\%)+Ward & 0.960 & 0.920 & 0.880 & 0.920 & 0.960 & 0.960 & 0.760 & 0.920 \\
CMDS(95\%)+K-means & 0.920 & 0.800 & 0.920 & 0.920 & 0.920 & 0.920 & 0.720 & 0.920 \\
CMDS(95\%)+Ward & 0.920 & 0.960 & 0.880 & 0.920 & 1.000 & 0.960 & 0.680 & 0.920 \\
CMDS(99\%)+K-means & 0.920 & 0.800 & 0.840 & 0.920 & 0.960 & 0.920 & 0.640 & 0.920 \\
CMDS(99\%)+Ward & 0.920 & 0.800 & 0.920 & 0.920 & 1.000 & 0.960 & 0.680 & 0.920 \\
\bottomrule
\end{tabular}
\end{table*}

\begin{table*}[t]
\centering
\caption{Leave-one-feature-out clustering ablation for matched JSD distances on the family benchmark. Entries report Adjusted Rand Index (ARI) across hierarchical clustering and CMDS-based clustering variants.}
\Label{tab:jsd_clustering_loo_ari}
\scriptsize
\setlength{\tabcolsep}{3pt}
\renewcommand{\arraystretch}{1.08}
\begin{tabular}{lcccccccc}
\toprule
\textbf{Algorithm} & \textbf{All Features} & \textbf{w/o Structural} & \textbf{w/o Mean Curvature} & \textbf{w/o Hydrogen Bond} & \textbf{w/o Electrostatics} & \textbf{w/o Hydrophobicity} & \textbf{Chemical Only} & \textbf{Geometry Only} \\
\midrule
Single & 0.300 & 0.282 & 0.282 & 0.282 & 0.335 & 0.461 & 0.282 & 0.649 \\
Complete & 0.556 & 0.495 & 0.248 & 0.493 & 0.722 & 0.357 & 0.286 & 0.756 \\
Average & 0.358 & 0.280 & 0.280 & 0.637 & 0.727 & 0.469 & 0.286 & 0.766 \\
Weighted & 0.358 & 0.280 & 0.286 & 0.580 & 0.760 & 0.469 & 0.286 & 0.766 \\
\midrule
CMDS(4D)+K-means & 0.476 & 0.235 & 0.216 & 0.456 & 0.494 & 0.357 & 0.235 & 0.802 \\
CMDS(4D)+Ward & 0.451 & 0.342 & 0.262 & 0.451 & 0.568 & 0.357 & 0.235 & 0.802 \\
CMDS(5D)+K-means & 0.571 & 0.224 & 0.261 & 0.565 & 0.727 & 0.396 & 0.248 & 0.787 \\
CMDS(5D)+Ward & 0.571 & 0.268 & 0.453 & 0.556 & 0.727 & 0.357 & 0.319 & 0.787 \\
CMDS(6D)+K-means & 0.565 & 0.248 & 0.504 & 0.392 & 0.722 & 0.521 & 0.288 & 0.752 \\
CMDS(6D)+Ward & 0.495 & 0.495 & 0.504 & 0.634 & 0.727 & 0.660 & 0.220 & 0.752 \\
CMDS(8D)+K-means & 0.392 & 0.300 & 0.294 & 0.392 & 0.717 & 0.521 & 0.255 & 0.773 \\
CMDS(8D)+Ward & 0.357 & 0.248 & 0.248 & 0.580 & 0.727 & 0.521 & 0.314 & 0.773 \\
CMDS(10D)+K-means & 0.502 & 0.300 & 0.294 & 0.555 & 0.717 & 0.357 & 0.255 & 0.777 \\
CMDS(10D)+Ward & 0.502 & 0.380 & 0.248 & 0.599 & 0.798 & 0.521 & 0.314 & 0.766 \\
CMDS(15D)+K-means & 0.357 & 0.308 & 0.357 & 0.555 & 0.722 & 0.357 & 0.308 & 0.777 \\
CMDS(15D)+Ward & 0.502 & 0.320 & 0.404 & 0.599 & 0.798 & 0.521 & 0.261 & 0.766 \\
CMDS(20D)+K-means & 0.412 & 0.249 & 0.357 & 0.392 & 0.798 & 0.484 & 0.261 & 0.766 \\
CMDS(20D)+Ward & 0.357 & 0.248 & 0.248 & 0.503 & 0.717 & 0.521 & 0.248 & 0.787 \\
\midrule
CMDS(80\%)+K-means & 0.309 & 0.275 & 0.275 & 0.565 & 0.389 & 0.248 & 0.268 & 0.773 \\
CMDS(80\%)+Ward & 0.309 & 0.248 & 0.255 & 0.556 & 0.439 & 0.289 & 0.220 & 0.773 \\
CMDS(90\%)+K-means & 0.392 & 0.300 & 0.504 & 0.555 & 0.727 & 0.357 & 0.248 & 0.737 \\
CMDS(90\%)+Ward & 0.357 & 0.380 & 0.504 & 0.599 & 0.727 & 0.521 & 0.319 & 0.787 \\
CMDS(95\%)+K-means & 0.357 & 0.300 & 0.357 & 0.555 & 0.773 & 0.456 & 0.255 & 0.766 \\
CMDS(95\%)+Ward & 0.502 & 0.380 & 0.248 & 0.599 & 0.809 & 0.521 & 0.314 & 0.787 \\
CMDS(99\%)+K-means & 0.369 & 0.249 & 0.357 & 0.392 & 0.798 & 0.484 & 0.261 & 0.766 \\
CMDS(99\%)+Ward & 0.357 & 0.248 & 0.248 & 0.599 & 0.717 & 0.521 & 0.248 & 0.787 \\
\bottomrule
\end{tabular}
\end{table*}

\begin{table*}[t]
\centering
\caption{Leave-one-feature-out clustering ablation for matched JSD distances on the family benchmark. Entries report cluster purity across hierarchical clustering and CMDS-based clustering variants.}
\Label{tab:jsd_clustering_loo_purity}
\scriptsize
\setlength{\tabcolsep}{3pt}
\renewcommand{\arraystretch}{1.08}
\begin{tabular}{lcccccccc}
\toprule
\textbf{Algorithm} & \textbf{All Features} & \textbf{w/o Structural} & \textbf{w/o Mean Curvature} & \textbf{w/o Hydrogen Bond} & \textbf{w/o Electrostatics} & \textbf{w/o Hydrophobicity} & \textbf{Chemical Only} & \textbf{Geometry Only} \\
\midrule
Single & 0.640 & 0.640 & 0.640 & 0.640 & 0.680 & 0.760 & 0.640 & 0.720 \\
Complete & 0.720 & 0.680 & 0.640 & 0.800 & 0.800 & 0.720 & 0.640 & 0.840 \\
Average & 0.680 & 0.640 & 0.640 & 0.800 & 0.800 & 0.720 & 0.640 & 0.840 \\
Weighted & 0.680 & 0.640 & 0.640 & 0.760 & 0.840 & 0.720 & 0.640 & 0.840 \\
\midrule
CMDS(4D)+K-means & 0.720 & 0.640 & 0.640 & 0.760 & 0.800 & 0.720 & 0.640 & 0.880 \\
CMDS(4D)+Ward & 0.720 & 0.680 & 0.680 & 0.720 & 0.840 & 0.720 & 0.640 & 0.880 \\
CMDS(5D)+K-means & 0.760 & 0.640 & 0.640 & 0.720 & 0.800 & 0.720 & 0.640 & 0.840 \\
CMDS(5D)+Ward & 0.760 & 0.680 & 0.720 & 0.720 & 0.800 & 0.720 & 0.680 & 0.840 \\
CMDS(6D)+K-means & 0.720 & 0.640 & 0.680 & 0.720 & 0.800 & 0.800 & 0.640 & 0.800 \\
CMDS(6D)+Ward & 0.680 & 0.680 & 0.680 & 0.800 & 0.800 & 0.800 & 0.640 & 0.800 \\
CMDS(8D)+K-means & 0.720 & 0.680 & 0.680 & 0.720 & 0.800 & 0.800 & 0.640 & 0.800 \\
CMDS(8D)+Ward & 0.680 & 0.640 & 0.640 & 0.760 & 0.800 & 0.800 & 0.680 & 0.800 \\
CMDS(10D)+K-means & 0.800 & 0.680 & 0.680 & 0.720 & 0.800 & 0.720 & 0.640 & 0.840 \\
CMDS(10D)+Ward & 0.800 & 0.760 & 0.640 & 0.760 & 0.840 & 0.800 & 0.680 & 0.840 \\
CMDS(15D)+K-means & 0.680 & 0.640 & 0.680 & 0.720 & 0.800 & 0.720 & 0.640 & 0.840 \\
CMDS(15D)+Ward & 0.800 & 0.680 & 0.720 & 0.760 & 0.840 & 0.800 & 0.640 & 0.840 \\
CMDS(20D)+K-means & 0.720 & 0.640 & 0.680 & 0.720 & 0.840 & 0.800 & 0.640 & 0.840 \\
CMDS(20D)+Ward & 0.680 & 0.640 & 0.640 & 0.800 & 0.800 & 0.800 & 0.640 & 0.840 \\
\midrule
CMDS(80\%)+K-means & 0.680 & 0.640 & 0.640 & 0.720 & 0.760 & 0.640 & 0.640 & 0.800 \\
CMDS(80\%)+Ward & 0.680 & 0.640 & 0.640 & 0.720 & 0.800 & 0.680 & 0.600 & 0.800 \\
CMDS(90\%)+K-means & 0.720 & 0.680 & 0.680 & 0.720 & 0.800 & 0.720 & 0.640 & 0.800 \\
CMDS(90\%)+Ward & 0.680 & 0.640 & 0.680 & 0.760 & 0.800 & 0.800 & 0.680 & 0.840 \\
CMDS(95\%)+K-means & 0.680 & 0.680 & 0.680 & 0.720 & 0.840 & 0.760 & 0.640 & 0.840 \\
CMDS(95\%)+Ward & 0.800 & 0.760 & 0.640 & 0.760 & 0.840 & 0.800 & 0.680 & 0.840 \\
CMDS(99\%)+K-means & 0.680 & 0.640 & 0.680 & 0.720 & 0.840 & 0.800 & 0.640 & 0.840 \\
CMDS(99\%)+Ward & 0.680 & 0.640 & 0.640 & 0.760 & 0.800 & 0.800 & 0.640 & 0.840 \\
\bottomrule
\end{tabular}
\end{table*}

\begin{table}[ht]
\centering
\caption{Directed threshold success percentages.}
\Label{tab:directed_threshold_success}
\begin{tabular}{rrrrr}
\toprule
Threshold & Relaxed Top-1 & Relaxed Top-2 & Strict Top-1 & Strict Top-2 \\
\midrule
0.30 & 100.00 & 100.00 & 100.00 & 100.00 \\
0.40 & 100.00 & 100.00 & 98.21 & 100.00 \\
0.50 & 94.64 & 96.43 & 83.93 & 89.29 \\
0.60 & 80.36 & 83.93 & 73.21 & 80.36 \\
0.70 & 75.00 & 78.57 & 66.07 & 69.64 \\
0.80 & 57.14 & 67.86 & 37.50 & 55.36 \\
0.90 & 33.93 & 48.21 & 16.07 & 30.36 \\
\bottomrule
\end{tabular}
\end{table}

\begin{table}[ht]
\centering
\caption{Bidirectional threshold success percentages.}
\Label{tab:bidirectional_threshold_success}
\begin{tabular}{rrrrr}
\toprule
Threshold & Strict Top-1 & Strict Top-2 & Relaxed Top-1 & Relaxed Top-2 \\
\midrule
0.30 & 100.00 & 100.00 & 100.00 & 100.00 \\
0.40 & 100.00 & 100.00 & 100.00 & 100.00 \\
0.50 & 100.00 & 100.00 & 100.00 & 100.00 \\
0.60 & 78.57 & 92.86 & 96.43 & 100.00 \\
0.70 & 53.57 & 67.86 & 75.00 & 85.71 \\
0.80 & 32.14 & 46.43 & 50.00 & 50.00 \\
0.90 & 7.14 & 21.43 & 25.00 & 42.86 \\
\bottomrule
\end{tabular}
\end{table}

\end{document}